\documentclass[10pt,letterpaper]{article}

\usepackage{amsmath}
\usepackage{amssymb}

\usepackage{amsmath}

\usepackage{graphicx}

\usepackage{cite}

\usepackage{caption}

\usepackage{color}

\usepackage{setspace} 
\usepackage[labelfont=bf,labelsep=period,justification=raggedright]{caption}

\makeatletter
\renewcommand{\@biblabel}[1]{\quad#1.}
\makeatother

\date{}

\begin{document}

\begin{flushleft}
{\Large
\textbf{Information flow through a model of the \textit{C. elegans} klinotaxis circuit
}
}
\\
Eduardo J. Izquierdo$^{\ast}$, 
Paul L. Williams, 
Randall D. Beer
\\
Cognitive Science Program, Indiana University, Bloomington
\\
$^\ast$ E-mail: edizquie@indiana.edu
\end{flushleft}

\section*{Abstract}

Understanding how information about external stimuli is transformed into behavior is one of the central goals of neuroscience.  Here we characterize the information flow through a complete sensorimotor circuit: from stimulus, to sensory neurons, to interneurons, to motor neurons, to muscles, to motion.  Specifically, we apply a recently developed framework for quantifying information flow to a previously published ensemble of models of salt klinotaxis in the nematode worm {\it Caenorhabditis elegans}.  The models are grounded in the neuroanatomy and currently known neurophysiology of the worm.  The unknown model parameters were optimized to reproduce the worm's behavior.  Information flow analysis reveals several key principles underlying how the models operate: (1) Interneuron class AIY is responsible for integrating information about positive and negative changes in concentration, and exhibits a strong left/right information asymmetry. (2) Gap junctions play a crucial role in the transfer of information responsible for the information symmetry observed in interneuron class AIZ. (3) Neck motor neuron class SMB implements an information gating mechanism that underlies the circuit's state-dependent response. (4) The neck carries more information about small changes in concentration than about large ones, and more information about positive changes in concentration than about negative ones. Thus, not all directions of movement are equally informative for the worm. Each of these findings corresponds to an experimental prediction that could potentially be tested in the worm.  Knowing the results of these experiments would greatly refine our understanding of the neural circuit underlying klinotaxis. Information flow analysis also allows us to explore how information flow relates to underlying electrophysiology.  Despite large variations in the neural parameters of individual circuits, the overall information flow architecture circuit is remarkably consistent across the ensemble, suggesting that information flow analysis captures general principles of operation for the klinotaxis circuit.

\newpage
\section*{Author Summary}

Perhaps one of the most ambitious goals of neuroscience is to understand how information about the environment is transformed by the nervous system to produce behavior. What information does each neuron have about the relevant stimuli? How is information from different components integrated?  How does that information change over time as it is transferred across different components of the circuit?  Although there have been many recent advances in applying information theory to the analysis of nervous systems, such analyses have usually been constrained, due to theoretical and experimental limitations, to specific areas of the brain, typically the sensory regions.  Here we apply recent theoretical advances in quantifying information dynamics to analyze a computational model of the neural circuit underlying salt klinotaxis behavior in the nematode worm {\it Caenorhabditis elegans}.  Our analysis provides the first characterization of information flow through a complete neuroanatomical circuit: from external stimuli, to sensory neurons, to interneurons, to motor neurons, to muscles, to motion. By analyzing information flow in an ensemble of model circuits, our analysis reveals that the information flow architectures for individual circuits share a number of general properties. 

\newpage
\section*{Introduction}
\label{sec:Introduction}

One of the grand challenges in neuroscience is to understand how an organism's behavior arises from the dynamical interaction between its brain, its body and its environment.  An important component of that challenge involves characterizing the flow and transformation of information through a complete neural circuit, from environmental stimuli, through sensory cells, through multiple recurrent layers of interneurons and motor neurons, and finally through muscles to produce behavior.  Information theory~\cite{Shannon1948,Thomas2006} has become an increasingly essential tool in neuroscience, with applications ranging from studies of neural coding~\cite{Rieke1999,Dayan2001} and the statistical structure of environmental stimuli~\cite{Attneave1954,Barlow1961}, to developing maps of functional connectivity in nervous systems~\cite{Friston2011,Sporns2011,Lizier2011,Wibral2011,Wibral2013}. However, there has not yet been an attempt to analyze the information flow through an entire sensorimotor circuit underlying a particular behavior. The obstacles to such an endeavor are both theoretical and experimental. 

On the theoretical side, the primary challenge is how to track the dynamic flow of information. Our approach to information flow analysis incorporates several recent extensions to the basic framework of information theory~\cite{Williams2010B,Beer2014}. First, motivated by the observation that standard information measures average across all measurement outcomes, we utilize measures of specific information to unroll such averages in order to quantify the information that components of the neural circuit provide about each specific value of the external stimulus, providing a more fine-grained analysis of informational relationships~\cite{DeWeese1999,Butts2003,Balduzzi2008,Borst1999,Eckhorn1975,Kjaer1994,Theunissen1991,Butts2006}. 
Second, we apply measures of dynamic information to track how information is gained and lost by individual components and transferred between components of the neural circuit over time\cite{Lizier2008,Lizier2012,Neymotin2011,Williams2010A}. Third and finally, we treat each component of the circuit as a random process, or time-indexed sequence of random variables, and characterize how information is carried by individual components and transferred between components over time. This extension is motivated by the observation that standard information measures, and even dynamic information measures like transfer entropy~\cite{Schreiber2000,Kaiser2002,Wibral2014}, are typically treated as atemporal; for example, transfer entropy is used to quantify the information that one random process $X$ at time $t$ transfers to a second random process $Y$ at time $t+1$ by averaging over all time indices. Instead, we unroll these averages to perform a more fine-grained analysis of the temporal structure of information flow (for a related approach to characterizing the temporal structure of information flow, see~\cite{Lizier2008,Lizier2010,Lizier2013}).

Experimentally, there are two primary challenges. First, complete  sensorimotor circuits underlying particular behaviors are rarely known. In part, this challenge can be addressed by focusing on simpler invertebrates. The nematode worm {\it Caenorhabditis elegans}, which has one of the simplest, most consistent and well-studied nervous systems, is uniquely qualified in this regard. The complete wiring diagram of its nervous system is known~\cite{White1986,Chen2006}, it is very well-characterized genetically~\cite{Brenner1974,Hobert2004}, it exhibits a rich behavioral repertoire~\cite{Sengupta2009,Hobert2003,Bargmann1993}, and putative circuits for several of these behaviors have been identified~\cite{deBono2005}. However, electrophysiological analysis is difficult in {\it C. elegans} due to its small size and pressurized body (although recent progress has been made using both optogenetic and whole-cell patch clamp techniques~\cite{Goodman1998,Zhang2007,Faumont2006,Stirman2011,Leifer2011}). To address this shortcoming in known neurophysiology, we utilize optimization techniques to generate an ensemble of electrophysiological parameter sets for model neural circuits that are consistent with the known connectivity and produce accurate klinotaxis behavior~\cite{Izquierdo2013}. Since such models can be rerun any number of times under varying conditions and all relevant variables can be simultaneously recorded, they also address the second major experimental challenge of information flow analysis: collecting the magnitude of data necessary to accurately estimate all of the time-varying information measures.

To demonstrate the utility of this approach, we focus here on salt klinotaxis, a form of chemotaxis in {\it C. elegans}. Klinotaxis involves gradual changes in orientation directed towards the source~\cite{Iino2009}, and provides a particularly interesting behavior to analyze for two main reasons. First, although the nematode detects sensory increases and decreases in concentration in different cells~\cite{Suzuki2008}, the circuit must combine information about the full spectrum of changes in concentration in order to steer gradually towards the source. Second, klinotaxis requires state-dependence: the circuit must respond to changes in concentration differently depending on the direction of its head swing and body posture, thus combining information from the environment with its own internal state to produce an appropriate response~\cite{Izquierdo2010}.  {\it C. elegans} also exhibits klinotaxis using sensory stimuli other than taste, including temperature~\cite{Luo2006}, odors~\cite{Yoshida2012}, and electric fields~\cite{Gabel2007}. Other species, such as the larvae of  {\it Drosophila melanogaster}, are also known to utilize klinotaxis for spatial orientation~\cite{GomezMarin2011,Kane2013}. A similar strategy has also been studied in humans following a scent trail while constrained to sample the environment through a single point using a nasal prism~\cite{Porter2007}. Thus, klinotaxis may be a representative example of spatial navigation.

The informational analysis is performed on a previously published model of the salt klinotaxis circuit in {\it C. elegans}~\cite{Izquierdo2013}.  The model is grounded in the available neuroanatomical~\cite{White1986,Chen2006}, neurophysiological~\cite{Suzuki2008}, and behavioral data~\cite{Iino2009}.  
The structure of the circuit was identified by exploring all the paths connecting the main salt chemosensory class, ASE~\cite{Bargmann1991}, and the neck motor class involved in modulating the amplitude of the sinusoidal locomotion, SMB~\cite{Gray2005}, while constraining the path length and the number of synaptic contacts of the resulting network until arriving at the minimal circuit connecting all sensory cells to all motor cells~\cite{Izquierdo2013}. The circuit involves four neuron classes: ASE, AIY, AIZ, and SMB; and the chemical synapses and gap junctions between them~(Figure 1). 
Where available, knowledge of the cellular physiology was incorporated into the model. Left and right ASE chemosensory neurons were modeled as ON and OFF cells, respectively, based on neurophysiological experiments~\cite{Suzuki2008}. 
The remaining unknown electrophysiological parameters, including the sign and strength of the connections, were optimized by running a large set of evolutionary searches aiming to reproduce the worm's behavior.  The majority of searches successfully produced circuits capable of klinotaxis behavior consistent with what has been observed in the nematode~\cite{Iino2009}. The result of the evolutionary searches was an ensemble of computational models of klinotaxis, which were analyzed through a combination of parameter study and dynamical systems analysis in a previous paper~\cite{Izquierdo2013}.  Recently, through a combination of laser ablations, calcium imaging, microfluidic devices, and optogenetic stimulation, experimental work has validated a number of important aspects of that model~\cite{McCormick2013}, including: (a) that the set of neurons in our minimal klinotaxis circuit are necessary; (b) that ASE neurons are active within the timescale of individual head swings, an important assumption in our model of the chemosensory neurons (additionally, evidence for this has also been found in closely related sensory neurons in {\it C. elegans}~\cite{Kato2014}); and (c) a left/right asymmetry in the circuit, which was first identified in an analysis of the parameters of our evolved networks~\cite{Izquierdo2013}, and which we further study now informationally.  

\begin{figure}[ht!]
\begin{center}
\includegraphics[width=0.45\textwidth]{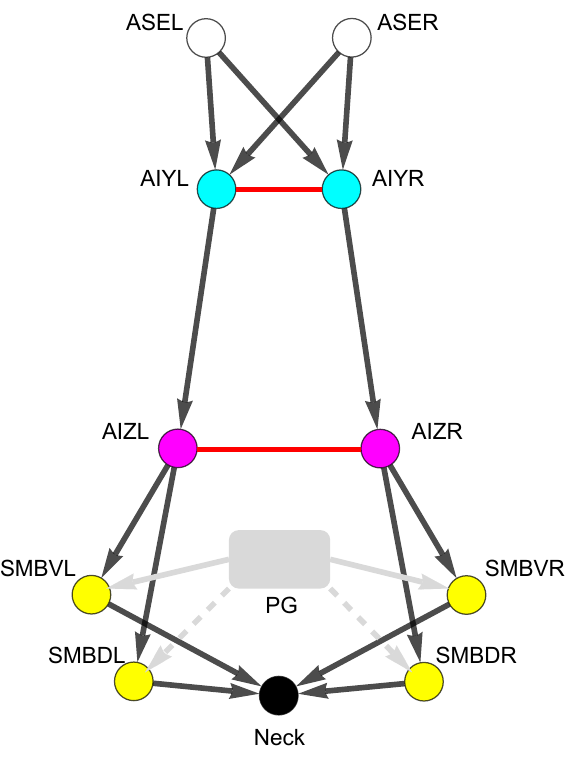}
\caption{
{\bf 
%Figure 1. 
Putative minimal {\it C. elegans} klinotaxis circuit (adapted from~\cite{Izquierdo2013}).}  
Chemosensory class, ASE (white).  Interneuron classes: AIY (blue) and AIZ (red).  Neck motor neuron class: SMB (yellow).  Neck angle (black).  All classes have left and right cells.  Motor neurons have additional dorsal and ventral pairs of cells.  Chemical synapses shown as black arrows.  Gap junctions shown as red undirected connections.  Motor neurons receive an oscillatory input from a pattern generator (gray). The pattern could be generated either through proprioceptive feedback or from a central pattern generator (see Methods). The oscillatory input is antiphase for ventral~(solid) and dorsal~(dashed) motor neurons.
}
\label{Fig1}
\end{center}
\end{figure}

An information flow analysis of a simple but complete model sensorimotor circuit allows us to engage two important theoretical issues in neuroscience. 
The first issue is understanding the relationship between the information flow and the underlying electrophysiology of the circuit. 
Generating time series recordings from the activity of cells in intact organisms is becoming increasingly common under a number of different conditions. Remarkably, in {\it C. elegans}  this includes whole-animal imaging~\cite{Prevedel2014} and imaging during freely-moving behavior~\cite{Clark2007,Shipley2014}. 
Such time series data is exactly what is required for information theoretic analysis. And thus, such analysis is likely to become increasingly common. However, despite progress in mapping the {\it C. elegans}  connectome~\cite{White1986}, a characterization of the relevant biophysical properties, including voltage-gated channels, synaptic properties, and neuromodulators, lags far behind.  Because we have an ensemble of models of the klinotaxis circuit with known parameters, and from which we can easily generate time-series recordings under any condition, we can explore the relationship between information flow and our mechanistic understanding of the models derived from knowledge of their neurophysiological parameters.

Even if complete biophysical knowledge about the nervous system were available, the problem of understanding the general principles by which it operates would remain unaddressed. This is the second theoretical issue that an information flow analysis of a complete sensorimotor model allows us to engage. 
In addition to the accelerated acquisition of time series data from cellular recordings, there is also considerable effort being invested in characterizing the biophysical properties of the nervous systems of a range of model organisms. 
What remains unclear is how to combine these large and diverse data to arrive at principles for how an organism generates any one specific behavior from the dynamical interaction between its brain, its body and its environment.  
This is particularly challenging in light of the individual variability observed in circuit properties~\cite{Gardner1989,Beer1999,Prinz2004}. 
Because the focus of information flow analysis is on the relationship between the activity of the circuit over time and certain behaviorally-relevant features of the environment, information flow analysis has the potential to uncover higher-level descriptions of the circuit's operation.
Furthermore, because we have access to an ensemble of model circuits, we can explore the possibility that information flow analysis captures general properties of the operation of many different klinotaxis circuits, despite substantial variations in the physiological properties of those circuits. 

In addition to these two broader issues, information flow analysis also allows us to address a number of specific questions about the neural basis of {\it C. elegans} klinotaxis. For example, where does the circuit integrate information about positive and negative changes in concentration to produce a unified action? What role do the gap junctions play in the operation of the circuit? How does the circuit combine information from the environment with its own internal state to steer in the correct direction? As information flows through the circuit, information about certain changes in concentration are preserved better than others. What specific information does the worm use to steer? 
We proceed in three phases.  We first examine the overall flow of information about changes in salt concentration through the best model klinotaxis circuit in our ensemble. We then analyze in detail each layer of the circuit, considering the specific information carried by each cell and the pathways along which information is transferred between cells. Finally, we look broadly at the similarities and differences between the best klinotaxis circuit and the rest of the model ensemble. 

\section*{Results}
\label{sec:Introduction}

\subsection*{Overall Flow of Mutual Information}
\label{sec:Overall}

As a foundation for the more detailed analysis to follow, we begin with a brief overview of the overall flow of information about changes in concentration through the highest-performing klinotaxis circuit. We performed the informational analysis using two different assays. 
The first assay, called the {\it concentration step assay}, involves giving the circuit a step in concentration at a specific time, where the magnitude of the step, $\Delta c$, is assumed to be distributed uniformly over the allowable range.   This method is common experimentally~\cite{Goodman1998,Suzuki2008,Thiele2009,Oda2011}.  
Second, we used an assay called the {\it information clamp assay}, which involves giving the circuit a constant change in concentration for the entire stimulus duration, where the magnitude of the change, $\dot{c}$, is assumed to be distributed uniformly over the allowable range. This is called the information clamp assay because ASE responds to changes in concentration~\cite{Suzuki2008,Thiele2009}, so a constant change in concentration means that ASE produces a constant output. Thus, we can think of the information that ASE carries as being clamped to a particular fixed value. This is also true for the components downstream from ASE that are independent of the sinusoidal input (AIY, AIZ). Crucially, the information clamp assay allows us to explore the specific role played by the sinusoidal input in regulating the flow of information in the SMB neck motor neurons. Within limits, this assay would also be experimentally feasible. The same two assays are used throughout the paper and are described in more detail in the Methods section.

We start with the overall flow of information about the magnitude of a step change in concentration, $\Delta c$~(Figure 2). Specifically, we calculate the time-varying mutual information $I(\Delta c;e(t))$  for each element $e$ of the system assuming a uniform distribution over $\Delta c$. We proceed by layers~(Figure 1), from the chemosensory neurons (ASEL/ASER), through the two layers of interneurons (AIYL/AIYR and AIZL/AIZR), to the motor neurons (SMBDL/SMBDR/SMBVL/SMBVR) and ultimately to the neck~($\theta$).

\begin{figure}[ht!]
\begin{center}
\includegraphics[width=0.8\textwidth]{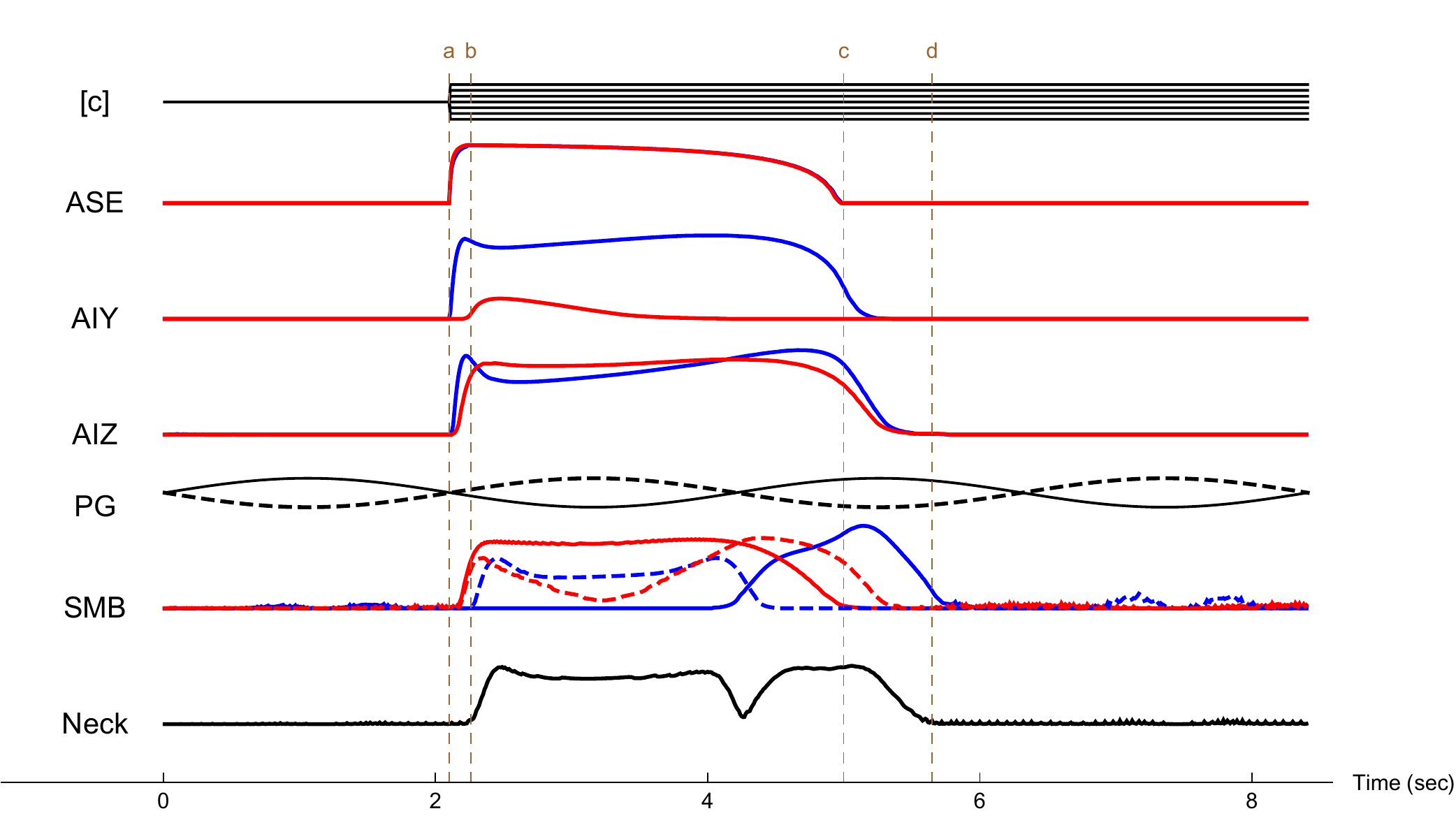}
\caption{
\noindent {\bf 
%Figure 2. 
Overall flow of mutual information for concentration step assay.} 
The top row illustrates the input to the circuit, salt concentration [c]. Mutual information over time, $I(\Delta c;e(t))$, for the left~(blue) and right~(red) cells are shown for each neuron class: ASE, AIY, AIZ, SMB, and the neck. Note ASEL's mutual information is the same as ASER's mutual information, so the red trace overlaps the blue trace. Ventral~(solid) and dorsal~(dashed) motor neurons receive anti phase inputs from an oscillatory pattern generator, PG. The dashed brown lines illustrate: onset of stimulus~(a), onset of neck response~(b), end of sensor response to stimulus~(c), and end of neck response~(d). Traces are shown for two full locomotion cycles~(8.2 secs).
}
\label{Fig2}
\end{center}
\end{figure}

The ASE chemosensory neurons detect $\Delta c$ directly. The time course of the response in the chemosensory neurons is determined by the rise and decay parameters (see Methods). Although ASEL is only sensitive to positive values of $\Delta c$ and ASER is only sensitive to negative values of $\Delta c$, their responses are otherwise identical: The rise in mutual information is sharp, and the information remains relatively stable for over half the locomotion cycle, after which there is a slow decay (ASE,~Figure 2). Thus, despite their functional specialization, the ASE responses, viewed at the coarse level of overall mutual information, are indistinguishable. 

The information response profiles for the interneurons are more interesting. Unlike for the ASE sensory neurons, the AIY interneurons exhibit a left/right asymmetry in the information that they carry about $\Delta c$ (AIY,~Figure 2). This is somewhat surprising since both AIY receive chemical synapses from both ASE neurons, which carry the same amount of information, and thus one might expect the AIY neurons to carry the same information. However, this is clearly not the case. In AIYL, significant information appears very rapidly and persists for the entire duration of its availability in ASE. In contrast, the much smaller quantity of information in AIYR appears later and vanishes quickly. Even more surprising, given the information asymmetry in AIY, is that information is symmetric at the next level in AIZ, with AIZL and AIZR exhibiting similar information profiles (AIZ,~Figure 2). Both of these informational features -- asymmetry in AIY and symmetry in AIZ -- turn out to be quite common across the ensemble of high-performing klinotaxis circuits. An important goal for our detailed analysis in the next section will be to examine the distinct informational pathways underlying these features.

Two additional factors complicate the informational analysis of the motor neurons (SMB,~Figure 2). First, unlike the sensory neurons and interneurons, the SMB receive a sinusoidal input representing the pattern generator that drives the undulatory wave of locomotion (PG,~Figure 2). Second, there are four SMB cells arranged symmetrically about the worm's body: a dorsal/ventral pair on the left (SMBDL/SMBVL) and a dorsal/ventral pair on the right (SMBDR/SMBVR). Since the worm locomotes on its side, it is the dorsal/ventral distinction that is important here. Accordingly, the oscillatory signal that the dorsal motor neurons receive is anti-phase to the oscillatory signal that the ventral motor neurons receive. The information response profiles for the motor neurons are complex and phase-dependent, and will be examined in detail in the next section.

Ultimately, information from all four motor neurons is integrated by the body to produce a change in neck angle. Since these motor neurons are driven by the oscillatory pattern generator, we might expect the information that the neck angle carries about $\Delta c$ to exhibit oscillations as well. However, the information in the neck actually holds relatively constant at a high value throughout the response (Neck,~Figure 2), with one exception: the information drops to essentially zero briefly around the midpoint of the locomotion cycle. This dip in information is the result of assumptions built into our model, which are discussed within the context of the detailed analysis of the neck. 

The overview in~Figure 2 also highlights the various timescales of information flow within the circuit. As an approximation of the propagation time through the entire circuit, the delay from the change in concentration to the initial response of the neck is 0.16 secs ($a$ to $b$, vertical dashed lines,~Figure 2). Information about $\Delta c$ is available in the sensors for 1.9 secs ($a$ to $c$), but persists in the model worm for an additional 0.65 secs after the sensory response ends ($c$ to $d$), giving the approximate duration of state-dependence in the circuit. The total duration of the neck response is 3.4 secs ($b$ to $d$), which corresponds to about 3/4 of a locomotion cycle. These informational timescales play an important role in understanding how this embodied circuit achieves reliable orientation when situated in a chemical environment. 

Finally, by viewing the entire network as an information channel, we can examine how well information is preserved through each layer of processing. This is best done using the information clamp assay, in which we consider a constant change in concentration $\dot{c}$ and compute  $I(\dot{c};l(t))$ for the set of elements in layer $l$ assuming a uniform distribution over $\dot{c}$. Of course, the ASE layer has perfect information about $\dot{c}$ (gray,~Figure 3). The largest loss of information occurs at the first layer, with AIY still preserving 80.2\% of the available information in ASE (blue). Most of the information in AIY is preserved by the AIZ layer, which contains 76.3\% of the original information (red). Due to the oscillating pattern generator input, information in the SMB layer fluctuates between 58.2\% and 76.2\% of the original information (orange), while the information preserved in the neck angle fluctuates between 0.08\% and 58.6\% (black). Averaged over a locomotion cycle, the neck angle contains about 48.4\% of the original information. Functionally, the information in the neck guides behavior. Therefore, we can say that the system as a whole preserves about half of the information about $\dot{c}$ that was originally available from the sensors. 

\begin{figure}[ht!]
\begin{center}
\includegraphics[width=0.45\textwidth]{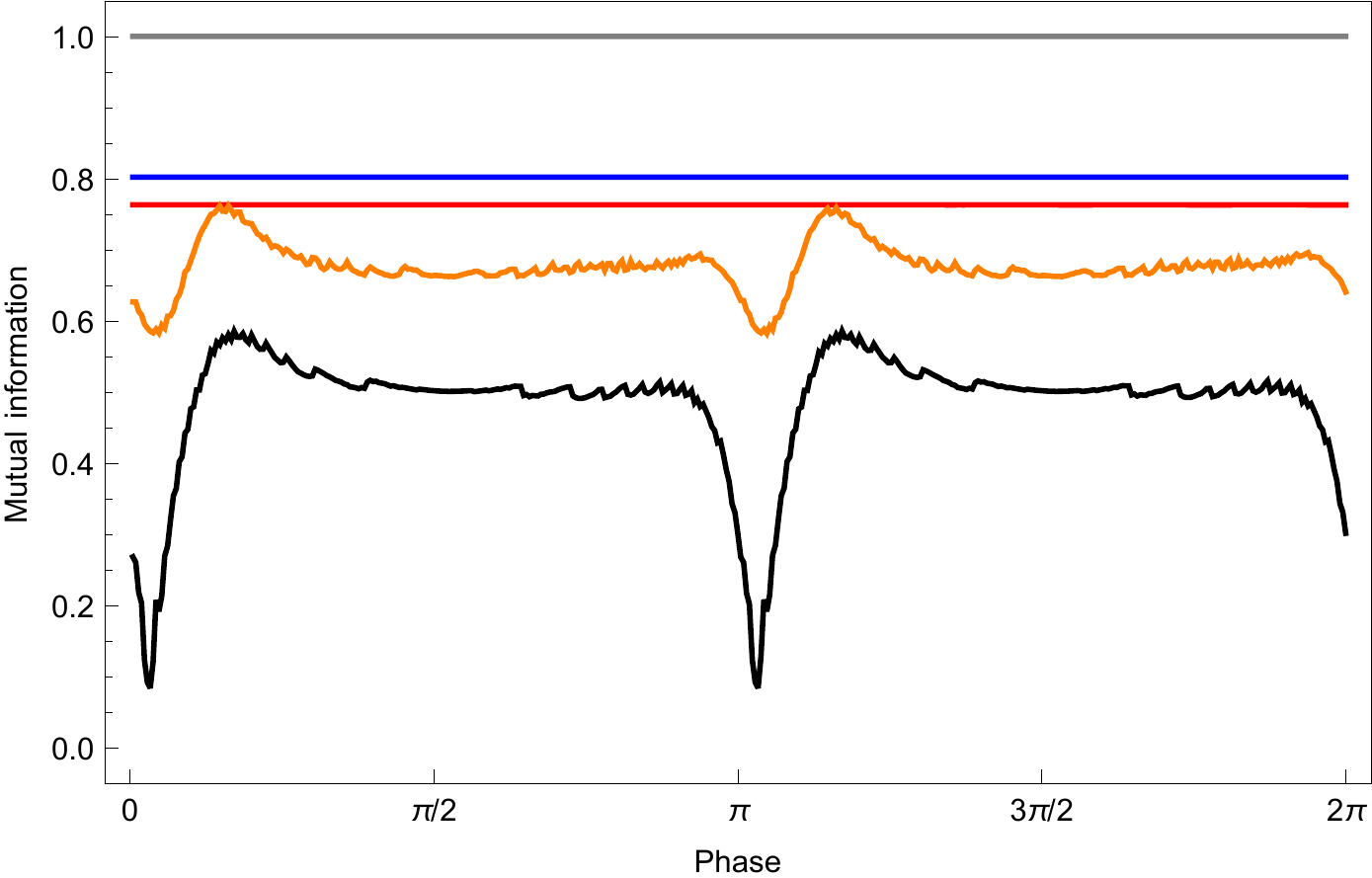}
\caption{
\noindent {\bf 
%Figure 3. 
Information preservation during information clamp assay.} 
Mutual information is shown for each of the classes in the network: ASE (gray), AIY (blue), AIZ (red), SMB (orange), and neck angle (black). Traces are shown for one cycle of locomotion.
}
\label{Fig3}
\end{center}
\end{figure}

\subsection*{Analysis by Neuron Class}
\label{sec:DetailedAnalysis}

Overall mutual information provides a broad foundation for understanding information flow in the network. In this section we analyze in more detail the information flow through each layer of the circuit. In particular, we study (a) the specific information that each neuron provides about particular changes in concentration; and (b) the pathways along which information is transferred from one neuron to another. All of the information theoretic quantities that we compute in this section are defined in the Methods section.

\subsubsection*{Information is Specialized in ASE}
\label{sec:ASE}

We begin with the ASE chemosensory neurons, which are modeled after physiological findings~\cite{Suzuki2008}: ASER and ASEL react to downsteps and upsteps in concentration, respectively. Since ASE responses are directly driven by changes in concentration, their informational analysis is straightforward. Nevertheless, this exercise serves to introduce the information-theoretic strategy and tools that we will use throughout and provides a baseline for our analysis of the remainder of the circuit.

First, we use specific information to unroll the mutual information in ASEL and ASER across different values for the change in concentration. Specifically, we compute the specific information that ASEL and ASER provide about $\Delta c$ as a function of the concentration step size $k$: $I(\Delta c = k;$ ASEL$)$ and $I(\Delta c = k;$ ASER$)$, respectively. Since ASEL and ASER detect only positive and negative concentration steps, respectively, it is not surprising that their specific information profiles reflect this pattern~(Figure 4). ASEL and ASER carry perfect information about the different ranges of $\Delta c$ values that each neuron detects (positive and negative values, respectively). Somewhat counterintuitively, each sensory neuron also provides a small amount of information about the stimulus range that it does not detect (smaller ridges in Figure 4). This is due to a kind of negative logic: for example, knowing that ASEL is off tell us that $\Delta c \le 0$, which rules out half of the possible stimulus values and thus provides one bit of information about the stimulus.  We call neurons such as these {\it informationally specialized} since they provide information about largely non-overlapping ranges of stimulus values.

\begin{figure}[ht!]
\begin{center}
\includegraphics[width=0.8\textwidth]{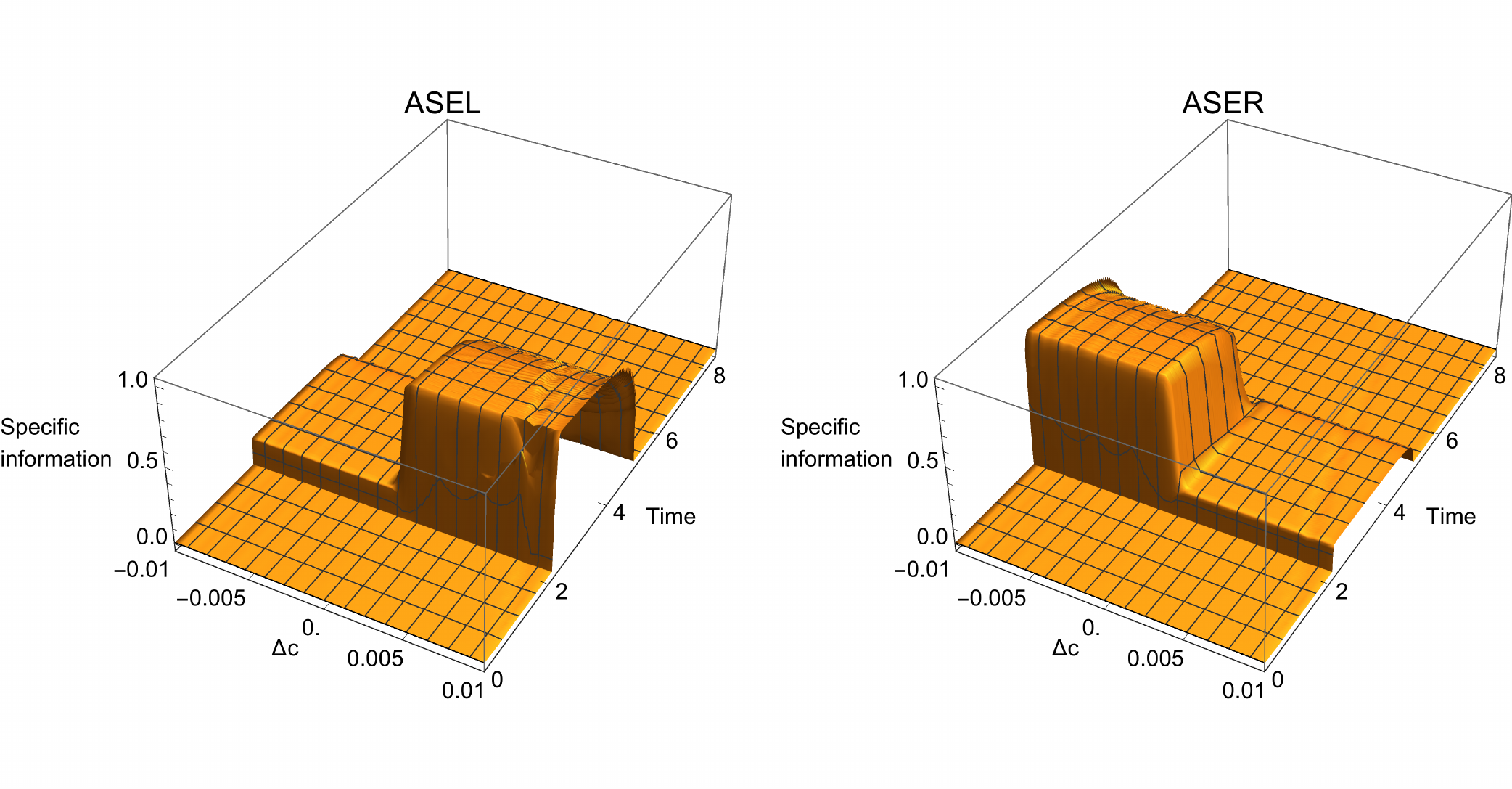}
\caption{
\noindent {\bf 
%Figure 4. 
Information analysis for ASE during concentration step assay.} Specific mutual information for left, $I(\Delta c=k;$ASEL$)$, and right, $I(\Delta c=k;$ASER$)$, cells over time.   
}
\label{Fig4}
\end{center}
\end{figure}

\subsubsection*{Information is Asymmetric in AIY}
\label{sec:AIY}

Unlike the ASE sensory neurons, the AIY cells exhibit a strong asymmetry in the amount of information that each cell carries about changes in concentration. This asymmetry can be seen quite clearly in plots of specific information for AIY~(Figure 5A). Although AIYL has more information about positive than negative steps, it maintains high information across the full range of $\Delta c$. In contrast, AIYR contains information only about the very largest positive steps~(Figure 5A). We call neurons such as these {\it informationally asymmetric} since the information carried by one dominates the information carried by the other across the full range of stimulus values.

\begin{figure}[ht!]
\begin{center}
\includegraphics[width=0.9\textwidth]{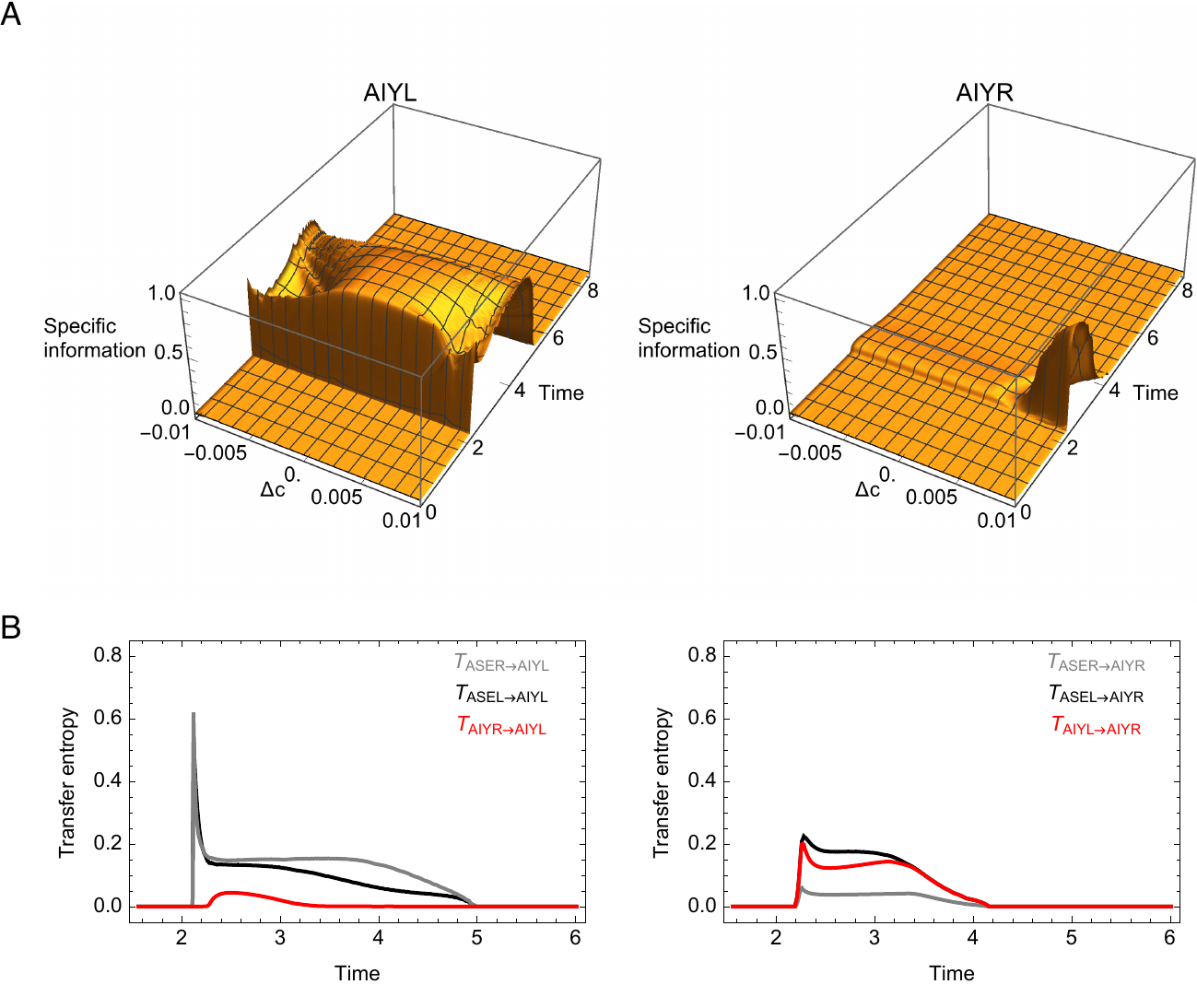}
\caption{
\noindent {\bf 
%Figure 5. 
Information analysis for AIY during concentration step assay.}   (A) Specific mutual information for left, $I(\Delta c=k;$AIYL$)$, and right, $I(\Delta c=k;$AIYR$)$, cells over time.  (B) Transfer entropy.  Black and gray traces represent the transfer entropy to AIY through the chemical synapses from ASER and ASEL, respectively.  Red traces represent the transfer entropy from the contralateral AIY cell through the gap junction. 
}
\label{Fig5}
\end{center}
\end{figure}

What is the origin of the informational asymmetry in the AIY layer?
There are two possible pathways of information flow into the AIY cells: (1) a direct route from ASE through the chemical synapses, and (2) an indirect route from the other AIY cell through the gap junction between them~(Figure 1).
Using transfer entropy, we can quantify the contributions of these different pathways~(Figure 5B). If we focus primarily around the time that the step occurs, we can see that both $T_{\textrm{ASEL} \rightarrow \textrm{AIYL}}$ and $T_{\textrm{ASER} \rightarrow \textrm{AIYL}}$ are high, whereas $T_{\textrm{AIYR} \rightarrow \textrm{AIYL}}$ is nearly zero. From this we conclude that the information in AIYL is transferred primarily through the chemical synapses via the direct route from ASE.  In contrast, the information in AIYR is transferred through both the chemical synapse from ASEL and across the gap junction with AIYL~(Figure 5B, right). The lack of information transfer across the chemical synapse from ASER to AIYR explains the lack of information about negative concentration changes in AIYR, since ASER is the only component upstream from AIYR that carries information about negative changes. There is overall less information transferred to AIYR than to AIYL, consistent with the overall asymmetry.

What explains the observed informational asymmetry in AIY? 
The saturating nonlinearity of the synaptic transfer function (see Methods) constrains the cell to respond to changes in concentration within a certain range. This range is a function of each cell's bias parameter in relation to the strength and sign of the incoming chemical synapses from the chemosensory neurons.
Therefore, to understand the observed asymmetry between the two AIY cells, we have to analyze the parameters of the cells that determine their dynamical behavior (Figure S1). 
The first thing to note is that network parameters were not constrained to be left/right symmetric during optimization. In other words, the optimization algorithm was allowed to produce solutions that met the behavioral criteria regardless of the specific parameter configurations. 
From our previous analysis~\cite{Izquierdo2013}, we know that there is one key parameter for the AIY cells that helps to explain their informational asymmetry.
Although the pair of incoming chemical synapses from ASEL and ASER are of similar strength and polarity for both AIY cells, the bias of the AIYR cell is far more negative than that of the AIYL cell, whose response range is nearly centered in the range of possible net input. The asymmetry is further maintained by a very weak gap junction between the AIY cells. Thus, although all parameters play a role in determining the ranges of sensitivity for these two cells, it is primarily the difference in the bias parameters for the two AIY cells and the weak gap junction between the AIY cells that accounts for the observed informational asymmetry.

\subsubsection*{Information is Symmetric in AIZ}
\label{sec:AIZ}

Despite the strong information asymmetry in the AIY layer, concentration step information becomes symmetric in the AIZ cells, as plots of specific information make clear~(Figure 6A). Although there is some variation across $\Delta c$, the specific information profiles of both AIZL and AIZR are similar to that of AIYL, with high amounts of information across the full range of $\Delta c$. We call neurons such as these {\it informationally symmetric} since they each carry approximately the same amount of information across the full range of stimulus values.

\begin{figure}[ht!]
\begin{center}
\includegraphics[width=0.9\textwidth]{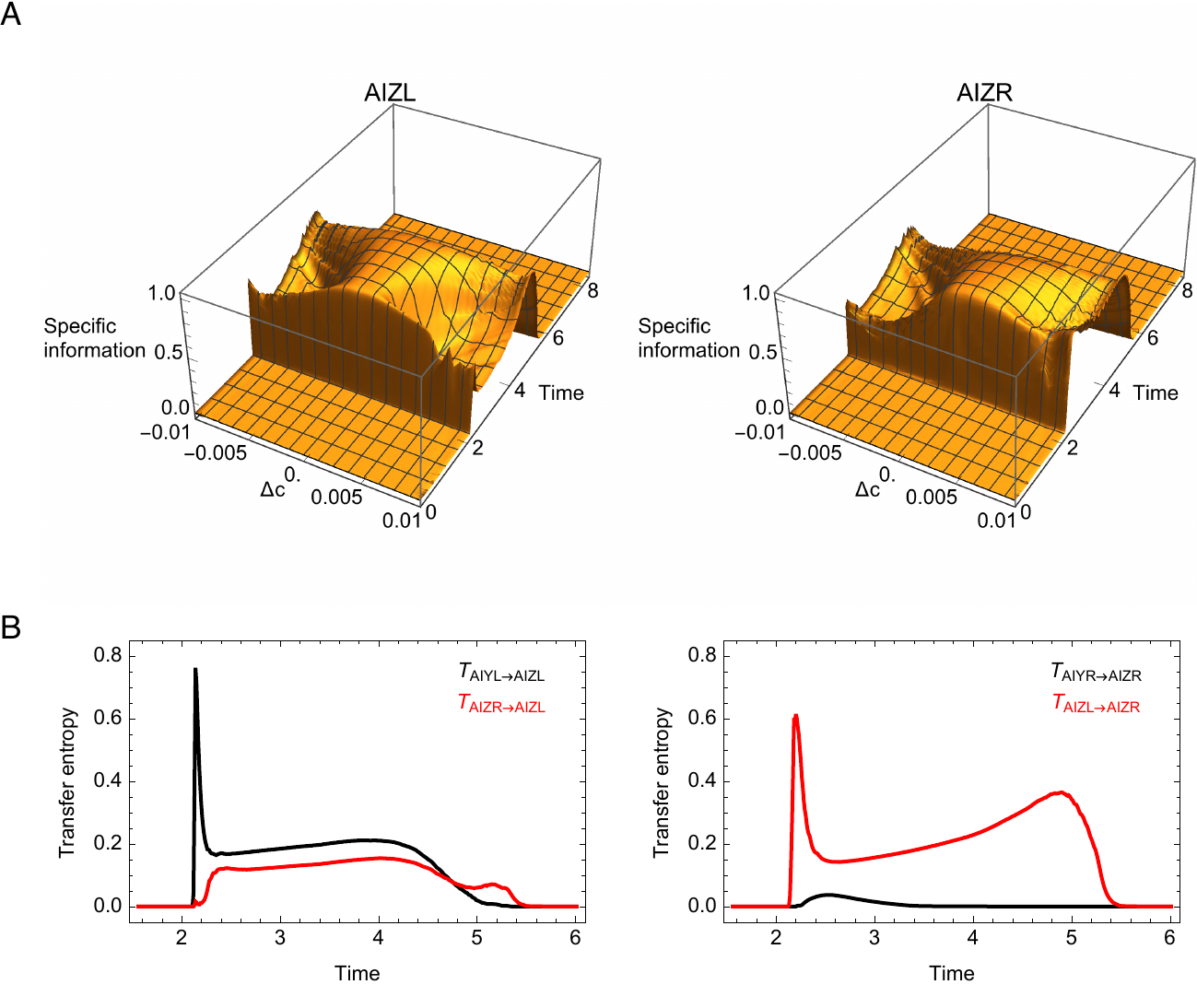}
\caption{
\noindent {\bf 
%Figure 6. 
Information analysis for AIZ during concentration step assay.}  (A) Specific mutual information for left, $I(\Delta c=k;$AIZL$)$, and right, $I(\Delta c=k;$AIZR$)$, cells over time. (B) Transfer entropy. Black trace represents the transfer of information from AIY to AIZ through the chemical synapse.  Red trace represents the transfer of information from the contralateral AIZ cell through the gap junction.
}
\label{Fig6}
\end{center}
\end{figure}

What is the origin of this informational symmetry in the AIZ layer? There are only two pathways for information to flow to an AIZ cell: (1) from the AIY neuron on the same side via a chemical synapse and (2) from the AIZ neuron on the opposite side via a gap junction~(Figure 1). If information was transferred only along the chemical synapses, we would expect the information in the left and right AIZ cells to be less than or equal to the information in the corresponding L/R AIY cells. This is the case for AIZL, but not for AIZR, which suggests that the gap junction connecting AIZL and AIZR is responsible for the information in AIZR. An analysis of transfer entropy confirms that, immediately after the step occurs, AIZL receives the majority of information directly through the chemical synapse, $T_{\textrm{AIYL} \rightarrow \textrm{AIZL}}$ (black trace,~Figure 6B). In contrast, AIZR receives its information not through the chemical synapse, $T_{\textrm{AIYR} \rightarrow \textrm{AIZR}}$, but rather through the gap junction, $T_{\textrm{AIZL} \rightarrow \textrm{AIZR}}$ (red trace,~Figure 6B). Thus, unlike in the AIY cells, the AIZ gap junction plays a major role in establishing the information profile in the AIZ layer.  

What explains the difference in information transfer between the AIY and the AIZ gap junctions? 
The gap junction connects two cells electrically so that current flows from the cell with more charge to the cell with less charge. As a result, the activation of the two cells tend to equalize. Once equalized, any change in activation in one of the cells, which under normal conditions is driven by external input, will cause a change in the activation of the other cell. Due to the nonlinearity of the synaptic transfer function, however, whether a change in net input will have an impact downstream depends on where the net input lies with respect to the sensitive region of the transfer function, which is determined by the cell's bias parameter. Therefore, the similarity of the bias parameter for the two cells influences their ability to exchange information. In the case of AIY  (Figure S1), AIYR responds to inputs from both the chemical synapses (red and blue region) and the gap junction, as the resting potentials are naturally equalized (dashed line). However, due to a large negative bias, the sensitive range of AIYR's synaptic transfer function lies too far to the right relative to the possible range of net input for changes in net input to result in changes in synaptic output. Consequently, no information is transferred to AIYR, neither through the chemical synapses from the sensory cells nor through the gap junction with AIYL.  
In the case of AIZ (Figure S2), AIZL receives information from the upstream cell through the chemical synapse, but AIZR does not.
However, once the cells equalize to their resting potentials, inputs to AIZL produce changes in both AIZL and in AIZR via the gap junction.  Since AIZR's sensitive region has a similar range as AIZL's, changes in activation propagated across the gap junction result in changes in the synaptic output of the cell, so that information is transferred effectively from AIZL to AIZR.  

\subsubsection*{Information Gating in SMB}
\label{sec:SMB}

Since the SMB neurons receive input from an oscillatory pattern generator in addition to the chemical synapses from the AIZ layer~(Figure 1), their response to a concentration step $\Delta c$ depends on the phase $\varphi$ of the oscillation when the step occurs. One way to visualize this phase dependence is to plot $I(\Delta c;$SMB$_i)$ for each SMB neuron as a function of $\varphi$~(Figure 7A). In order to simplify the plot, we show $I(\Delta c;$SMB$_i)$ only at a fixed delay of 50 msec after a step occurs, which corresponds to the time it takes the information in the AIZ neurons to stabilize after a step (AIZ,~Figure 2). Here we see that each SMB neuron acts as a kind of gate, allowing $\Delta c$ information from the AIZ layer to pass through at some phases, but blocking or strongly attenuating it at others. Of course, examining this effect at a single delay gives a very limited window into what is in fact a temporally-extended response. In order to visualize the cumulative effect of this response, we can average plots like~Figure 7A over all possible delays for a full locomotion cycle~(Figure 7B). Although the gating is seen less sharply in this case, the phase dependence of information transmission from the SMB layer is still quite clear.

\begin{figure}[ht!]
\begin{center}
\includegraphics[width=0.45\textwidth]{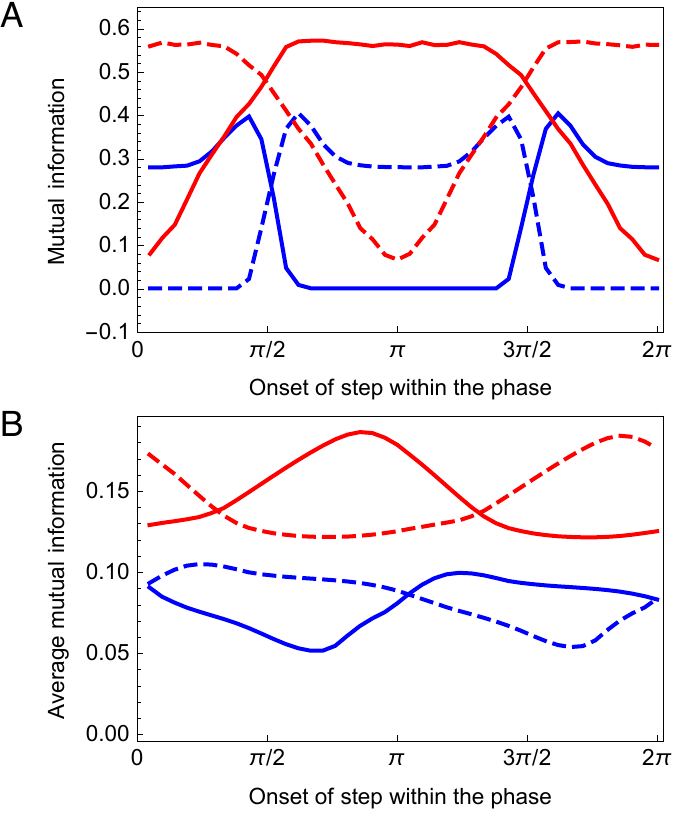}
\caption{
\noindent {\bf 
%Figure 7. 
Information gating in SMB motor neurons during concentration step assay.} (A) Mutual information for each SMB neuron as a function of the onset of the concentration step relative to the phase of locomotion. The mutual information is measured at a fixed delay of 50 msec after the step occurs. Left cells shown in blue. Right cells shown in red. Ventral~(solid) and dorsal~(dashed) traces. (B) Mutual information averaged over time as a function of the onset of the concentration step relative to the phase of locomotion.
}
\label{Fig7}
\end{center}
\end{figure}

Antiphase information gating in the dorsal and ventral motor neurons is responsible for the state-dependent response that allows the model worm to steer. 
A change in concentration received at different phases of locomotion produces identical chemosensory signals that are sent simultaneously to both the dorsal and ventral motor neurons, yet these signals have asymmetrical effects that allow the model worm to undergo either a dorsal or ventral turning bias, as appropriate for orienting correctly to the gradient.
For example, if an increase in concentration received during a dorsal-to-ventral sweep attenuates dorsal turning, then the same increase in concentration received during a ventral-to-dorsal sweep should instead attenuate ventral turning. 
By opening and closing the flow of information through the dorsal and ventral motor neurons in antiphase, the worm  generates different responses to the same stimuli depending on its phase of locomotion. 

In order to understand how information gating is implemented at the neuronal level, we need to consider the synaptic transfer functions for the motor neurons (Figure S3). 
From previous work~\cite{Izquierdo2010,Izquierdo2013}, we know that the sensitive region of the synaptic transfer functions are shifted relative to the range of the oscillatory input. Consequently, when one motor neuron is in its sensitive region, the other motor neuron is in its insensitive region, and vice versa. It is this alternating pattern of saturating nonlinearity in the SMB cells that explains the information gating: while the motor neuron in its sensitive region transfers information to the neck, the other neuron blocks the flow of information by saturating its output.

Finally, we examine how information about changes in concentration is distributed across the individual SMB cells. In order to more easily visualize the corresponding specific information, we make two simplifications. First, we switch to the information clamp assay in order to remove the phase dependence of the cell's response to the concentration step assay. Second, we consider the dorsal/ventral pairs jointly on each side. This simplification is motivated by the fact that, because the network parameters are dorsal/ventral symmetric (see Methods), the dorsal/ventral information profiles on each side are identical except for a phase shift, whereas the left/right profiles are quite different~(Figure 7B). The resulting plots of $I(\dot{c}=k;$SMBL$)$ and $I(\dot{c}=k;$SMBR$)$ are shown in~Figure 8.  Note that the $\dot{c}$ information in the right SMB motor neurons is strongly biased toward increases in concentration, whereas information in the left SMB motor neurons is strongly biased toward small decreases in concentration. However, despite this apparent specialization, SMBR carries more information than SMBL across the range of stimulus values. In the case of positive changes in concentration, SMBL carries no information, whereas SMBR carries some. In the case of negative changes in concentration, SMBR carries a similar amount of information as SMBL, so that a lot of the information about negative changes in SMBL is shared by SMBR.  Thus, most of the information carried by the SMB layer resides in the SMBR pair, with the SMBL pair making only small contributions at very specific points in time.

\begin{figure}[ht!]
\begin{center}
\includegraphics[width=0.8\textwidth]{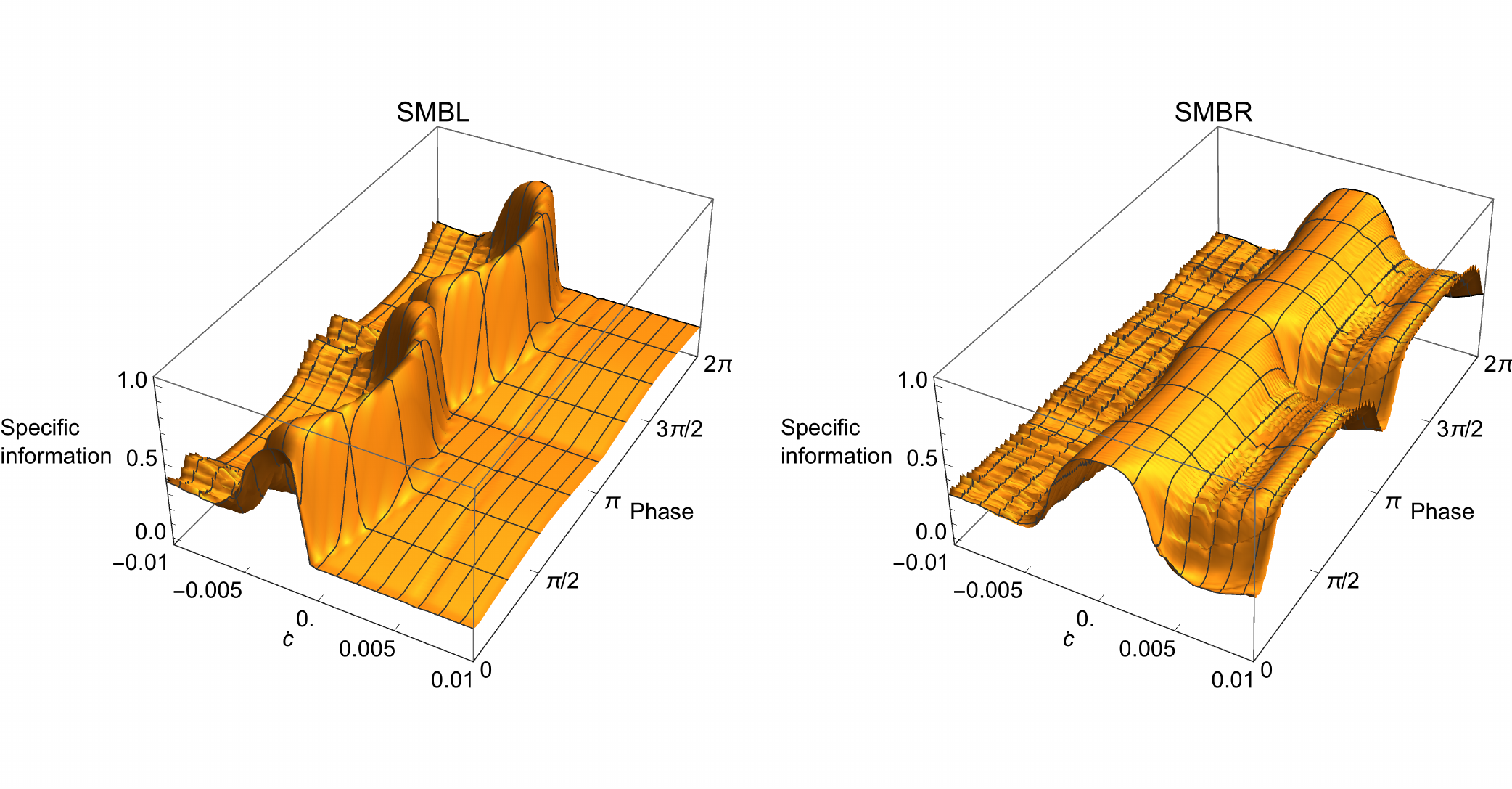}
\caption{
\noindent {\bf 
%Figure 8. 
Information analysis in left and right pairs of SMB cells during information clamp assay.}  Specific mutual information for left, $I(\dot{c}=k;$SMBL$)$, and right, $I(\dot{c}=k;$SMBR$)$, pairs over time. Surfaces shown for one cycle of locomotion.  
}
\label{Fig8}
\end{center}
\end{figure}

\subsection*{Functional Information in the Neck}
\label{sec:Neck}

The neck integrates information from the motor neurons, converting it to a neck angle that steers physical movement. Like the motor neurons, the information profile of the neck also varies across the locomotion cycle due to the oscillation of the pattern generator. Using the information clamp assay, we examine how information about concentration changes in the neck angle varies with both time and stimulus, with particular interest in the behavioral significance of these variations.

The specific information in the neck, $I(\dot{c}=k;\theta)$, is shown in~Figure 9A. There are two notable features of this plot. First, the $\dot{c}$ information available in the neck angle is remarkably consistent over time, remaining at a high value except for one sharp dip near the midpoint of the locomotion cycle. This brief drop in information is due to the assumption that the worm steers by modulating only the amplitude, and not the frequency, of the neck oscillation. Enforcing this assumption requires that the parameters of the SMB neurons are dorsal/ventral symmetric (see Methods), which in turn implies that SMB pairs receive the same inputs when the pattern generator crosses zero halfway through each locomotion cycle so that their inputs to the neck cancel out. Functionally, this corresponds to the transition between ventral and dorsal bending in the neck.

\begin{figure}[ht!]
\begin{center}
\includegraphics[width=0.9\textwidth]{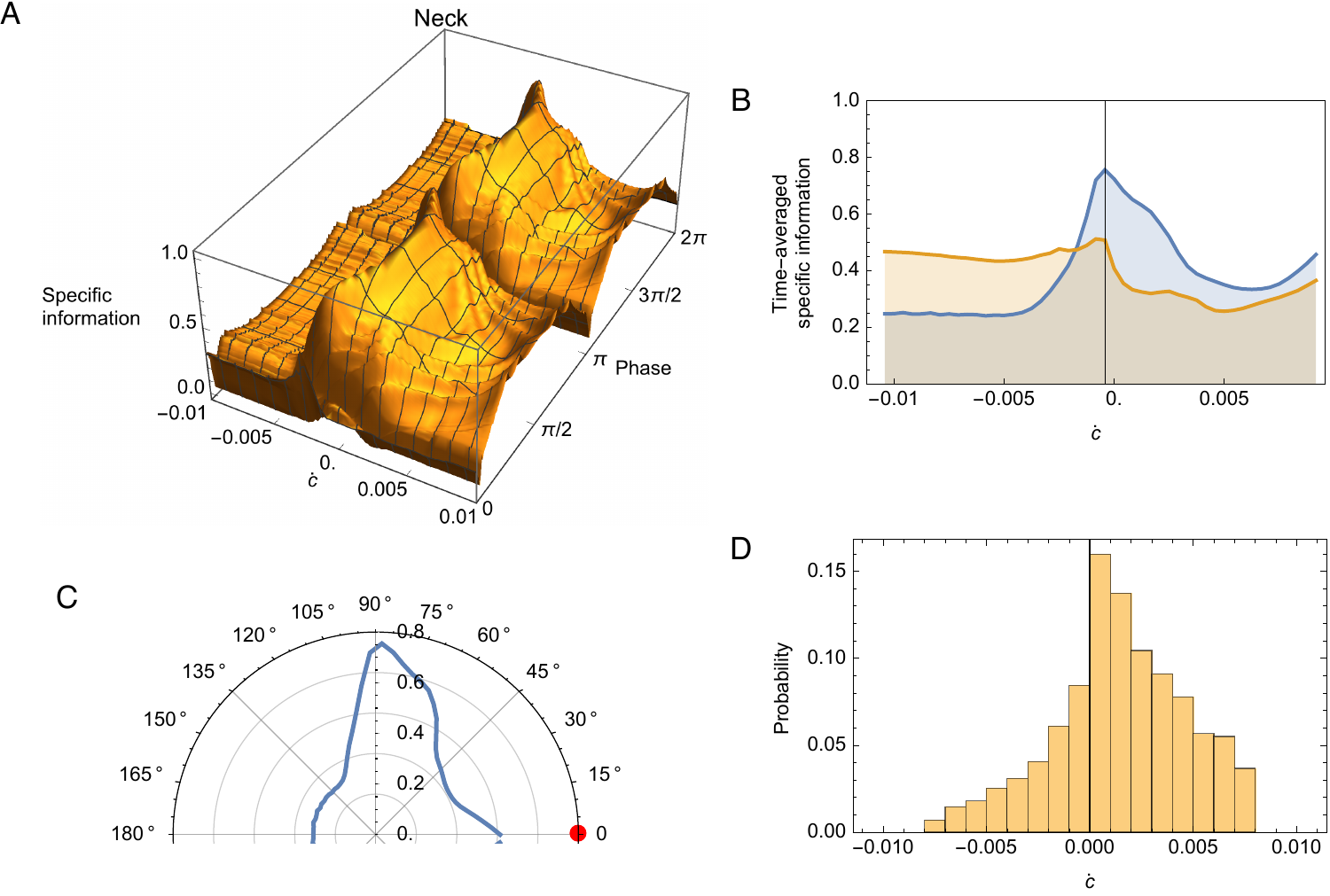}
\caption{
\noindent {\bf 
%Figure 9. 
Information analysis in the neck during information clamp assay.} (A) Specific information over time, $I(\dot{c}=k;\theta)$. Surface shown for one cycle of locomotion. (B) Specific information averaged over time as a function of the change in concentration, calculated using a sample of the stimulus feature chosen from a uniform distribution (blue) and from the model-generated empirical distribution shown in panel D (yellow). (C) Specific information averaged over time as a function of the worm's instantaneous orientation with respect to the implied peak of the gradient (red disk, 0 degrees). Data shown only from 0 to 180 degrees due to symmetry. (D) Empirical distribution of changes in concentration perceived by the simulated worm during a typical klinotaxis run.
} 
\label{Fig9}
\end{center}
\end{figure}

The second notable feature of specific information in the neck is its variation with $\dot{c}$. Given the consistency of this information over time, the variation is best visualized by averaging over a locomotion cycle~(blue trace, Figure 9B). Two properties immediately become clear from this plot. First, the neck has more information about small changes in concentration than about larger changes. Second, the neck carries more information about positive changes in concentration than about negative changes. In order to make sense of these results, we must place them within the context of the worm's behavior.  

In a freely moving worm, concentration changes result from the instantaneous movement of the worm's head relative to the peak of the gradient. Figure 9C shows the time-averaged specific information in the neck as a function of the orientation of the head with respect to the direction towards the gradient peak (red disk). Note that the circuit receives the most information when the head is moving perpendicular to the direction of the gradient peak (90 degrees). Note also that the circuit has more information about concentration changes when the head is moving in the direction of the peak (0 degrees), than when moving away from it (180 degrees). This indicates that not all directions of movement are equally informative for the worm; movements perpendicular to the peak of the gradient, which generate only small changes in concentration, turn out to be the most informative.

Interestingly, the distribution of concentration changes encountered by a freely-moving worm engaging in klinotaxis behavior~(Figure 9D) has a remarkably similar structure to the average specific information carried by the circuit~(Figure 9B), with both having peaks at $\dot{c} \approx 0$ and a bias toward positive changes in concentration. In other words, the freely moving worm spends the majority of each run moving perpendicular or towards the gradient peak, and the information extracted by the klinotaxis circuit~(Figure 9B) is well-tuned to the corresponding distribution of concentration changes encountered during klinotaxis behavior~(Figure 9C).  

In the absence of any experimental characterization of the actual distribution of concentration changes encountered by a worm performing klinotaxis, analyzing information flow under the assumption of a uniform distribution of sensory inputs is the best that one can do. However, we can use the distribution of concentration changes encountered by our model (Figure 9D) as a estimate of the corresponding distribution for the worm and then re-analyze the information flow using this predicted empirical distribution. When this is done, we observe that the specific information in the neck is more uniform across all stimulus values (yellow, Figure 9B), in contrast with the more highly skewed information with the uniform distribution (blue, Figure 9B). This seems consistent with the idea that the network has been optimized for the statistical structure of its environment. The specific information takes into account both how well a given stimulus is encoded and how ``surprising'' (i.e., improbable) it is, or equivalently, how much information we stand to gain by learning its value. Therefore, a more uniform distribution of specific information values with the empirical distribution suggests the neck carries more information about less surprising stimuli and less information about more surprising stimuli. A more rigorous test of this idea would require a computation of the optimal input distribution for a given klinotaxis circuit using, for example, the Blahut-Arimoto algorithm~\cite{Blahut1972,Arimoto1972}. This is an important direction for future work.

\subsection*{Information architecture}
\label{sec:IA}

Our detailed examination of the best-performing circuit allows us to characterize its general pattern of information flow, which we will refer to as its {\it information architecture}~(Figure 10B). 
The information architecture provides a static summary of the flow of information through the circuit, where the opacity of each node represents the amount of information that component carries averaged over time, the color of each node represents what specific information that component carries (where blue indicates information about positive concentration changes, and red indicates information about negative changes), and the arrows between nodes represent the amount of information transferred between them. Several of the previously identified features are apparent from this diagram. First, we see that information is specialized in the two ASE cells. It is then transferred through the chemical synapses and combined in AIYL. AIYR receives little to no information through either of the chemical synapses or the gap junction. The information in AIYL is transferred through the chemical synapse to AIZL, and is then transferred through the gap junction to AIZR.  Both AIZ cells transfer their information to the SMB cells downstream, although more is transferred from AIZR to SMBR. Finally, the SMB cells transfer information to the neck. Interestingly, we can contrast the information architecture with the underlying parameters of the circuit~(Figure 10A). For example, from the strengths of the incoming connections to AIY alone, we would expect information to be symmetrical at the AIY layer, rather than the observed asymmetry discussed previously within the context of the detailed analysis of AIY.  Also, from the strengths of the gap junction, we would expect transfer to be stronger between the AIY cells than between the AIZ cells, rather than the other way around as discussed previously within the context of the detailed analysis of AIZ. In general, the same neural parameters can give rise to many different information architectures depending on the surrounding body and environment, and the same information architecture can be produced by many different sets of neural parameters, so that knowledge of the structural and information flow architectures of a circuit provide complementary insight into the circuit's structure and function, respectively (for related ideas, see~\cite{Battaglia2014}). 

\begin{figure}[ht!]
\begin{center}
\includegraphics[width=0.8\textwidth]{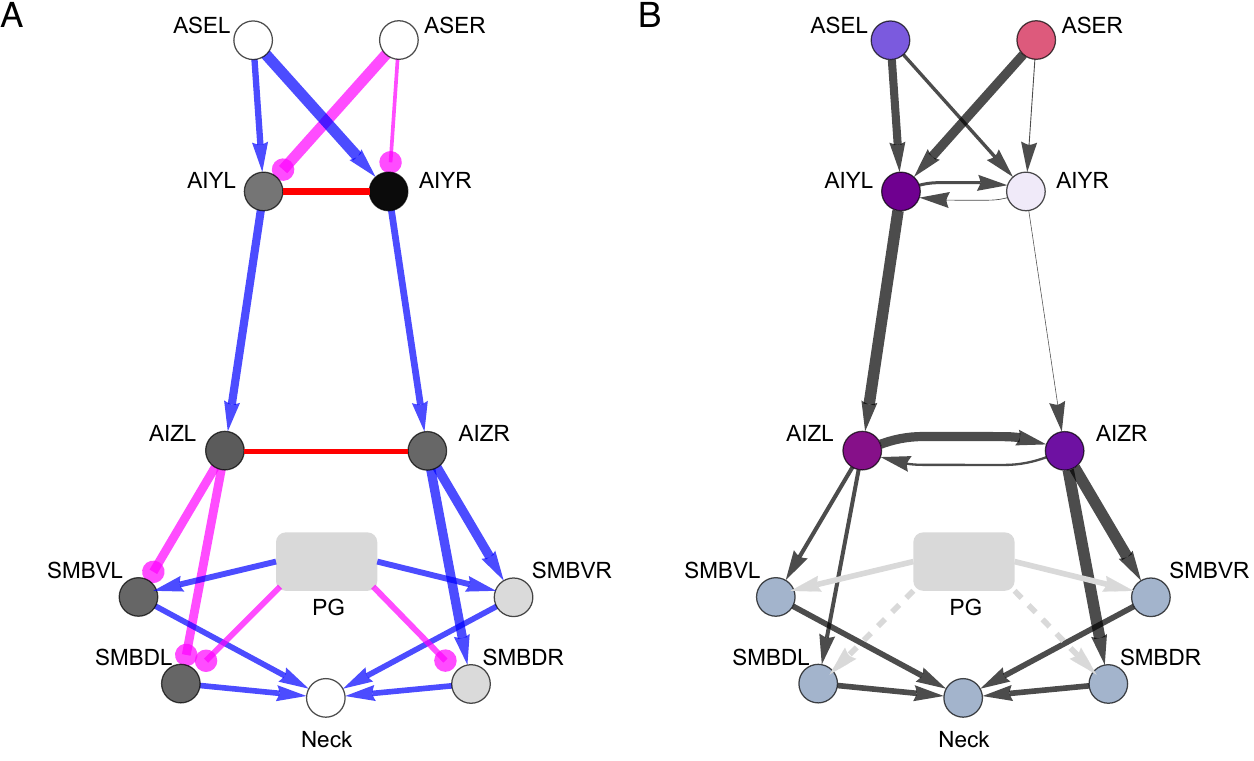}
\caption{
\noindent {\bf 
%Figure 10. 
Structure and function in the best evolved circuit. 
}
(A) Structural architecture. The strength of the chemical and electrical connections in the circuit are represented by the thickness of the lines connecting the nodes. Excitatory chemical synapses are shown in blue. Inhibitory chemical connections are shown in magenta. Gap junctions are shown in red.  Notice that unlike chemical synapses, gap junctions are undirected. The bias of the cells are represented by the shade of gray of the node. The bias determines how much activation the cell needs to change its output, and is only shown for cells AIY, AIZ, and SMB (see Methods). Dark cells require more activation to change their output. Lighter cells require less activation. 
(B) Information architecture.  The average amount of information in each cell is represented by the opacity of the node. The type of specific information in each cell is represented by the color of the node, and was calculated using the time-averaged specific information.  The proportion of blue/red indicates the amount of information the cell carries about positive/negative changes in concentration, respectively. The amount of information transferred between two cells is represented by the thickness of the arrows, and is calculated using the maximum transfer entropy. For ASE, AIY, and AIZ cells, we used the concentration step assay.  For SMB cells and the neck, we used the information clamp assay.
} 
\label{Fig10}
\end{center}
\end{figure}

\subsection*{Similar Information Architecture from Disparate Parameters}
\label{sec:Population}

To this point, we have examined in some detail the information flow of the best-performing circuit only.  However, the complete ensemble of klinotaxis circuits contains many others with comparable performance. Interestingly, the individual neuronal parameter values vary widely from circuit to circuit (Supplemental Figure 1). In this section, we examine the extent to which the information architecture of the best circuit is representative of this neurophysiologically diverse set of successful circuits. We focus on the same ensemble of highest-performers analyzed in previous work~\cite{Izquierdo2013}.  Specifically, we compare the information response profiles of the best circuit to the mean $\pm$ standard deviation response profiles of the ensemble.

The information profiles of the ASE, AIY and AIZ cells are shown in~Figure 11 for the concentration step assay. The most important thing to notice about these plots is the strong similarity between the profiles of the ensemble mean (solid trace) and those of the best circuit (dotted trace). First, the rise in $\Delta c$ information in ASE is always sharp and always persists for a little over half a locomotion cycle~(Figure 11A). The fall in information shows more variability, which implies that the sensory neurons have a wider range of values for the parameter $M$ (decay rate, see Methods and Figure S1). This suggests that there is less selection pressure on the decay rate ($M$) than on the rise rate ($N$) parameter. Second, only one of the AIY cells integrates $\Delta c$ information, consistently creating a strong left/right information asymmetry at this level of the circuit~(Figure 11B). Finally, $\Delta c$ information is consistently symmetric across the AIZ cells~(Figure 11C).

\begin{figure}[ht!]
\begin{center}
\includegraphics[width=0.45\textwidth]{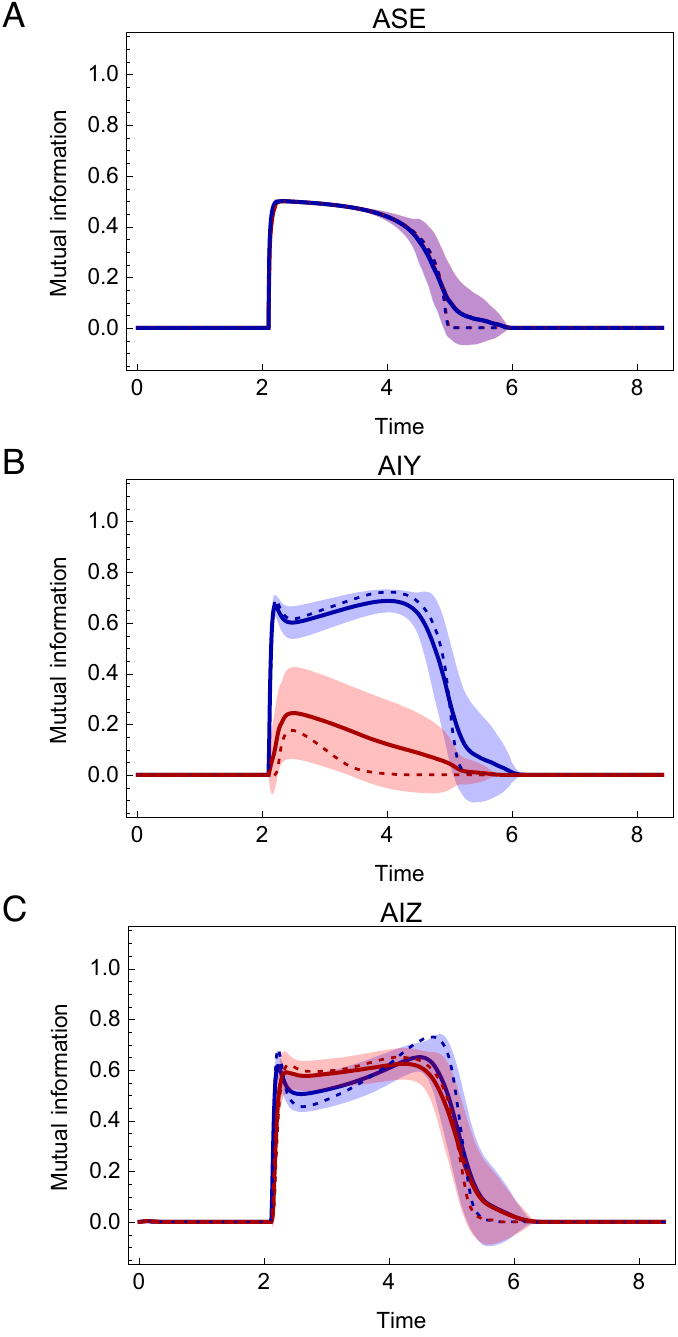}
\caption{
\noindent {\bf 
%Figure 11. 
Mutual information during concentration step assay in the population of successful circuits.} (A) Chemosensory neuron ASE: left (blue) and right (red) cell. (B) Interneuron AIY: cell with highest mutual information in blue, the other cell in red. (C) Interneuron AIZ: cell downstream from the AIY cell with highest mutual information in blue, the other cell in red. Mean (solid trace) and standard deviation (shaded area) for the ensemble of successful networks. Mutual information for the best circuit shown as a dotted trace.
}
\label{Fig11}
\end{center}
\end{figure}

The striking similarities between the best circuit and the ensemble average continue when we examine the pattern of information transfer between levels~(Figure 12). At the AIY layer, all of the $\Delta c$ information is consistently transferred into the primary AIY cell directly via the chemical synapses from ASE~(Figure 12A), with little use of the gap junction~(Figure 12C). In addition, at the AIZ layer, the $\Delta c$ information is first transferred from the primary AIY cell to the downstream AIZ cell via the chemical synapse~(Figure 12B), and only then transferred via the gap junction to the other AIZ cell~(Figure 12D). Thus, as in our analysis of the best circuit, we see that the gap junctions consistently play a major role in balancing information at the AIZ layer. 

\begin{figure}[ht!]
\begin{center}
\includegraphics[width=0.9\textwidth]{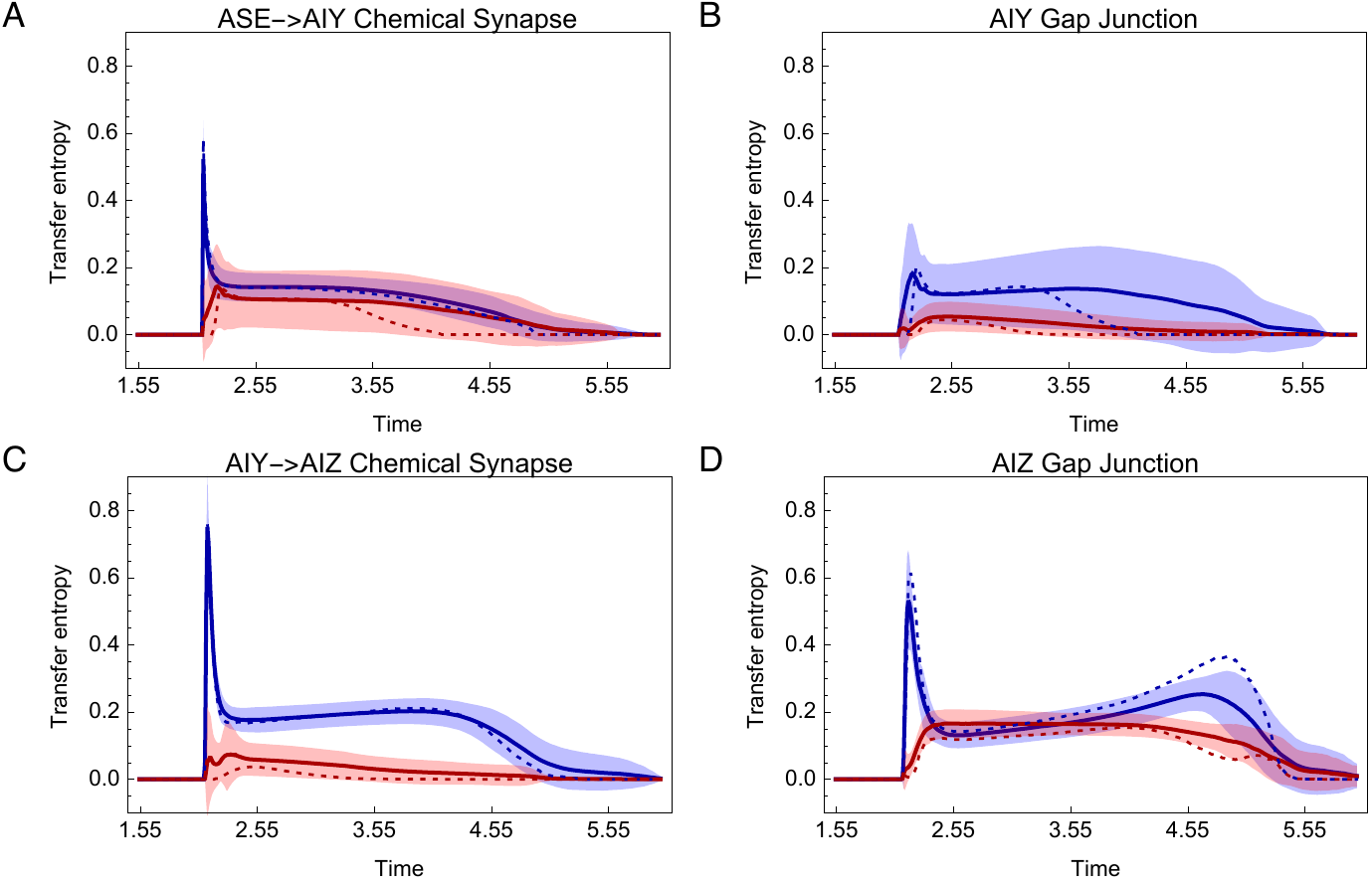}
\caption{
\noindent {\bf 
%Figure 12. 
Transfer entropy in the population of successful circuits.} (A) Through chemical synapses from ASE to: AIY cell with most information (blue), and to the AIY cell with lowest information (red). The traces show the mean (solid trace) and standard deviation (shaded area) for the chemical synapses from both ASE cells. (B) Through chemical synapse from the AIY cell with most information to the AIZ cell downstream (blue), and from the AIY cell with lowest information to the AIZ cell downstream (red). (C) Through the AIY gap junction: from the cell with most information to the cell with lowest information (blue), vice versa (red). (D) Through the AIZ gap junction: from the cell downstream of the AIY cell most information to the cell downstream of the AIY cell with lowest information (blue), vice versa (red). Mean (solid trace) and standard deviation (shaded area) for the ensemble of successful networks.  Transfer entropy for the best circuit shown as a dotted trace.
}
\label{Fig12}
\end{center}
\end{figure}

Despite greater variability in the information profiles for the SMB layer, there are still strong similarities between the best circuit and the ensemble average. For example, the concentration step assay reveals clear evidence of $\Delta c$ information gating at the SMB layer across the ensemble~(Figure 13A,B). In addition, as with the best circuit, an analysis of mutual information for the information clamp assay demonstrates a left/right asymmetry in SMB across the ensemble, with one side consistently providing more information about $\dot{c}$ than the other~(Figure 13C). However, we also see evidence that the best circuit is a bit of an outlier at the SMB level: the ensemble has more balanced information in the left/right pairs than does the best circuit.

\begin{figure}[ht!]
\begin{center}
\includegraphics[width=0.45\textwidth]{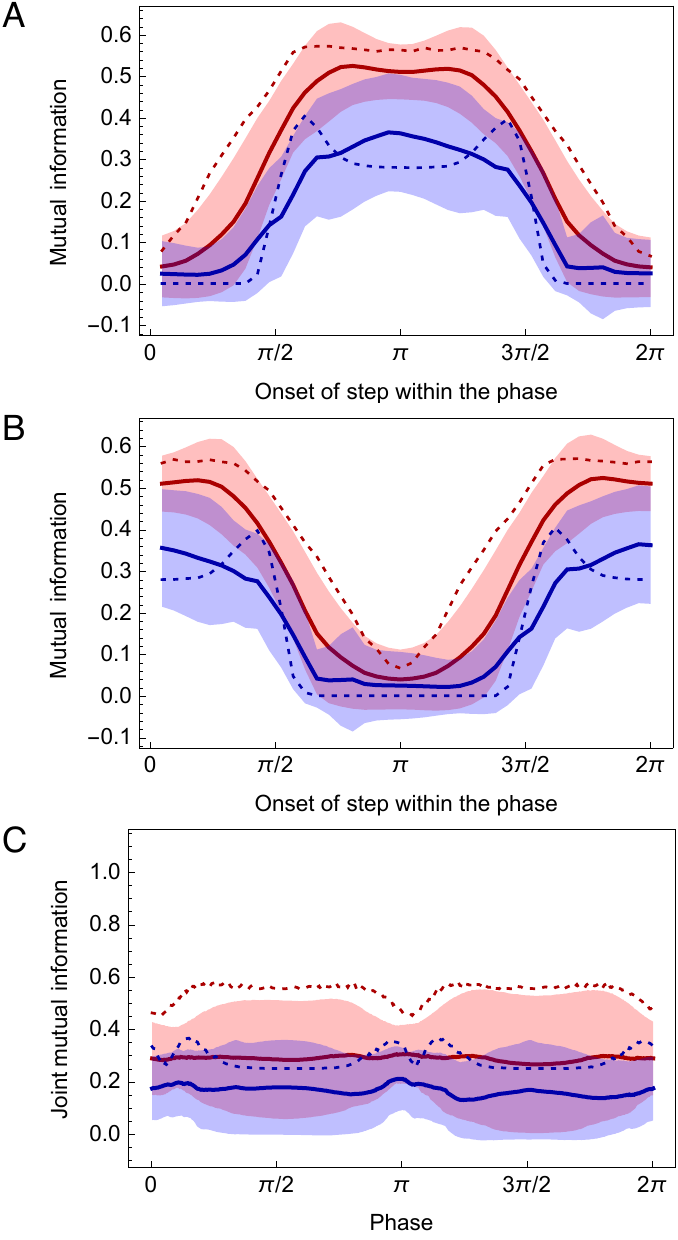}
\caption{
\noindent {\bf 
%Figure 13. 
Information gating in the population of successful circuits.} (A, B) Mutual information for each SMB neuron as a function of the phase of locomotion when the step in concentration is given, at a fixed delay of 50 msec after the step occurs. The four motor neurons are organized in two ways. The dorsal pair and ventral pair are categorized by their phase: Dorsal/ventral pairs with highest mutual information around $\pi$ of the locomotion phase (A), pairs with highest mutual information around $0/2\pi$ (B). The left and right motor neurons are categorized by the cumulative information they carry: The cell with highest information is shown in blue, the other cell is shown in red. Mean (solid trace) and standard deviation (shaded area) for the ensemble of successful networks. (C) Mutual information for the joint left and right pairs during information clamp assay. Pair with highest mutual information shown in red, other pair shown in blue. 
}
\label{Fig13}
\end{center}
\end{figure}

Finally, we consider information about $\dot{c}$ in the neck using the information clamp assay~(Figure 14). Despite the outlier status of the best circuit at the SMB layer, at the neck the ensemble mean information profiles are very close to those of the best circuit, with relatively small dispersion. Specifically, as in the best circuit, the mean amount of information about $\dot{c}$ across the ensemble remains relatively constant throughout the locomotion cycle, preserving about half of the original information available from the chemosensors~(Figure 14A). In addition, the  mean amount of specific information about particular values of $\dot{c}$ across the ensemble, averaging over one locomotion cycle, is peaked at $\dot{c}\approx 0$  and higher for increases in concentration than for decreases, as in the best circuit~(Figure 14B).

\begin{figure}[ht!]
\begin{center}
\includegraphics[width=0.45\textwidth]{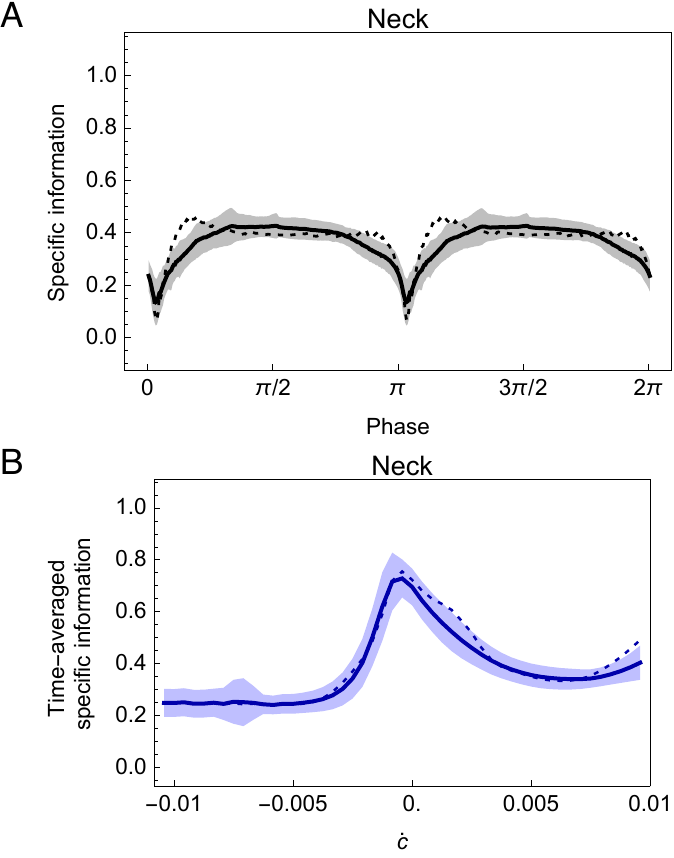}
\caption{
\noindent {\bf 
%Figure 14. 
Functional information in the neck during information clamp assay in the population of successful circuits.} (A) Mutual information over time shown for one cycle of locomotion. (B) Specific information averaged over time as a function of the change in concentration. Mean (solid trace) and standard deviation (shaded area) for the ensemble of successful networks. Best circuit (dotted trace).
}
\label{Fig14}
\end{center}
\end{figure}

We conclude from this analysis that the results obtained from the best klinotaxis circuit are in fact fairly representative of the entire ensemble of successful circuits (see Figure 10B). This is a somewhat surprising conclusion, given the large variations in synaptic strengths, intrinsic neuronal properties (Figure S4), and circuit dynamics across the ensemble~\cite{Izquierdo2010}. The consistency of the information flow architecture across the ensemble suggests that information flow analysis may be particularly well-suited for capturing general principles of operation that cut across the substantial variability and complex idiosyncrasies of individual klinotaxis circuits.

\section*{Discussion}
\label{sec:Discussion}

In this paper, we set out to analyze how information about changes in salt concentration flows through a putative minimal circuit for {\it C. elegans} klinotaxis~\cite{Izquierdo2013,McCormick2013}.  Our goal was to demonstrate how the tools of information theory can be used to characterize the flow of information throughout a complete sensorimotor circuit.  We proceeded in three stages.  First, we examined the overall flow of information by unrolling mutual information across time for each component of the highest-performing circuit.  Second, we analyzed in detail each layer of this circuit.  We unrolled information across stimulus values for each cell, and we quantified information transfer between individual circuit elements. Finally, we considered the similarities and differences between the best klinotaxis circuit and the rest of the model ensemble.  

Information flow analysis gave insight into specific questions about the neural basis of klinotaxis in {\it C. elegans}, centered around two issues that make this behavior particularly interesting. First, given the functional asymmetry of the ASE chemosensory cells~\cite{Suzuki2008}, the klinotaxis circuit has to integrate information about positive and negative changes in concentration. How is this integration achieved? Second, unlike klinokinesis~\cite{PierceShimomura1999}, the other chemotaxis strategy that the worm employs, klinotaxis requires state-dependence: the network has to combine information from the environment with its own internal state to produce an adequate response~\cite{Izquierdo2010}. How does this integration occur? 

Given that information about positive and negative changes in concentration is segregated in the chemosensory neurons (ASEL detects upsteps in salt concentration and ASER detects downsteps in salt concentration~\cite{Suzuki2008}), the klinotaxis circuit must ultimately integrate this information in order to steer. In principle, integration could occur at any level in the minimal circuit. However, given that each AIY cell receives connections from both ASE chemosensors~\cite{White1986}, the AIY layer is in the most favorable position. There are several different ways that integration could be achieved at the AIY level: (1) Both AIY cells could integrate concentration information via the chemical synapses from ASE; (2) One AIY cell could integrate the information via its chemical synapses and then pass it to the other via the gap junction between them; (3) Only one AIY cell could integrate the concentration information. Of these three possibilities, option (3) requires the fewest parameters to be tuned and it is this option that we consistently found in our model ensemble. Thus, our analysis makes a specific prediction about the information profile in AIY. Given the ubiquity of this profile in the ensemble, a failure of this prediction would suggest that the assumptions behind the minimal model need to be revisited. 

Strictly speaking, the integration of ASE information is only necessary if we assume that access to high amounts of information about the entire range of concentration changes is required for successful klinotaxis. Our analysis showed that, although each sensory neuron provides perfect information about only half of the range of concentration changes, each sensory neuron also provides a small amount of information (1 bit) about the other half of the range. This suggests that either chemosensory neuron alone could in principle drive at least a rudimentary form of the behavior if only a small amount of information about the entire range of concentration changes is required. For example, an ASEL-only circuit could only make graded adjustments when it was moving toward the peak, but it could still detect when it was moving away from the peak via the lack of activity in ASEL. Indeed, ablation studies have shown that an ASER-only circuit can successfully chemotax, but the results have been mixed for ASEL-only circuits~\cite{Iino2009,McCormick2013}. Simulated ablation studies in a previous model have also demonstrated successful chemotaxis behavior for ASEL-only and ASER-only conditions~\cite{Izquierdo2010}. However, in both experiments and simulations, the single-ASE behavior is quite different from normal klinotaxis and the success rate varies significantly with environmental circumstances.

Our analysis also provides insight into the mechanism by which the klinotaxis circuit combines information about concentration changes with information about the phase of its locomotion to steer in the right direction at the right time. We know from previous studies that the klinotaxis circuit must be state-dependent: the correct response to a given stimulus depends on the phase of the head swing when it is received~\cite{Izquierdo2010}.  For example, an upstep received during a dorsal-to-ventral swing should produce an increase in the ensuing dorsal turn, whereas an upstep received during a ventral-to-dorsal swing should produce a decrease in the same turn. Although our previous modeling studies suggested how saturating nonlinearity in the SMB cells might account for this state-dependence. Here we place this mechanism in a broader context as a kind of information gate, with the circuit using the oscillatory signal to alternately open and close the flow of concentration information through the motor neurons. Since this oscillatory signal is antiphase for dorsal and ventral pairs of motor neurons, the timing of the gates is also antiphase. This leads to different responses to the same upstep or downstep stimuli at any specific time, depending on which of the gates are open and which are closed. Although the information gating mechanism is crucial for the state-dependence in klinotaxis, the proposal that it occurs in the SMB neck motor neurons is a result of the model's assumption that the neck motor neurons receive antiphase oscillatory input from the worm's locomotion. In principle, the gating can occur in any component of the network that receives the anti phase oscillatory input generated during locomotion, including in the interneurons or even in the neck motor muscles themselves, where the antiphase oscillatory input could arise from postulated stretch-receptors~\cite{Boyle2012}.

In addition to specific questions relating to the neural basis of {\it C. elegans} salt klinotaxis, access to an ensemble of models whose parameters are fully known, and from which we can easily generate time-series recordings under any condition, allows us to explore the relationship between the information flow of a circuit and the mechanistic understanding that we can derive from knowledge of its parameters. 
Information flow analysis provided insight into crucial aspects of the operation of the circuit, albeit at a relatively high level of abstraction: AIY cells were informationally symmetric, AIZ cells were informationally asymmetric, and SMB cells acted as information gates in phase with the worm's locomotion. In order to answer questions about the mechanistic underpinnings of these features, we had to consider the specific parameters and the dynamics of the model.
Nevertheless, the suggestion is that the higher level of abstraction involved in  information flow analysis can be useful to focus more detailed analysis of the underlying parameters and dynamics of the model on only the most functionally relevant parts. 
This is a particularly relevant point given that technology for generating time-series recordings from intact biological systems during behavior is growing faster than the characterization of the underlying biophysical properties of their relevant circuits.  

Even if we had all of the details about the nervous system of a biological organism, understanding the principles underlying the neural basis of its behavior remains one of the main challenges in neuroscience.  
How do we combine the diverse and large amount of data needed to arrive at a set of principles for how the organism generates behavior from the dynamical interaction between its brain, its body and its environment?
By focusing on a higher level of abstraction, information flow analysis captures unique insights into the operation of the circuit in two interesting ways. First, it allows the analysis to focus on the functionally relevant aspects of the circuit. Second, it allows for a more manageable comparison between circuits that perform the same behavior, even if their biophysical properties are entirely different. 
This was particularly useful in the ensemble of klinotaxis circuits, where despite large variations in individual neurophysiological parameters, the analysis revealed a single information flow architecture that was unique among all high-performing circuits in the ensemble. 
Indeed, the uniqueness of the klinotaxis information architecture suggests an interesting general lesson for neuroscience: Perhaps the truly universal principles are to be found not among the neurophysiological details, which can vary substantially~\cite{Gardner1989,Beer1999,Prinz2004}, but instead at the more abstract level of patterns of information flow within a nervous system.

Despite being underdetermined by the available biological data, the uniqueness of the information flow architecture among the ensemble of circuits suggests that the constraints on which the model are based are reasonably strong. It would be very satisfying if the information flow architecture that we have identified turns out to be the one utilized by the worm. However, this architecture is only as good as the assumptions that went into the model. It is conceivable that, as the model assumptions are revised in light of new experimental data, key features of the information architecture will change. 
We mention briefly a few different directions for expanding the model.
First, there are several candidate neurons that would be particularly useful to include in next iterations of the model, including chemosensory neurons ADF and ASH~\cite{Thiele2009,McCormick2013} and interneuron classes AIA and AIB~\cite{Oda2011}, all of which have direct connections with one or more neurons in the minimal circuit~\cite{White1986}. 
Second, the specific location of the postulated information gating depends on which components of the circuit receive the oscillatory input. Recent evidence that specific subcellular regions of RIA carry information about dorsal and ventral head swings~\cite{Hendricks2012} is promising because it suggests a possible neuronal source of the oscillatory pattern. However, RIA is not directly presynaptic to any neuron in the minimal circuit~\cite{White1986}. This suggests that there may be either intermediate neurons involved, or an additional set of neck motor neurons involved (e.g., RIA is presynaptic to neck motor neurons SMD and RMD). 
Other components in the network are also viable alternatives to implement information gating, including the neck muscles themselves~\cite{Stretton1985}. 
Third, as we learn more about the neurophysiology of the underlying circuit, it may also become necessary to complicate the neural model that we currently employ. Recent advances in optogenetics in the freely-moving worm~\cite{Faumont2006}, as well as new experimental designs to study head swings in microfluidic devices~\cite{McCormick2011} will accelerate the characterization of the neurophysiology in the proposed circuit. 
Finally, we would like to test the neural mechanism for steering postulated in our work within the context of a biomechanical model of forward locomotion (e.g.,~\cite{Boyle2012}).  
 
Another important direction for future work is to continue to develop the tools of information dynamics. 
We mention briefly three different directions.
First, when analyzing the information in a system, it is necessary to consider not only the information carried by individual variables, but also the information that may be encoded redundantly or synergistically by multiple variables. 
Although the concepts of synergy and redundancy have been of great recent interest in several areas in neuroscience~\cite{Harder2013,Bertschinger2014}, the standard measures confound synergistic and redundant interactions and have problematic interpretations when more than three variables are involved~\cite{Brenner2000,Latham2005,Panzeri1999,Schneidman2003}. We, along with several other groups, are currently working to develop measures of synergy and redundancy that would overcome these problems~\cite{Williams2010A,Bertschinger2014,Griffith2014a,Griffith2014b,Harder2013}. 
Second, in this paper we studied information flow only under open-loop conditions, meaning that the worm's movement did not influence the concentration changes that it experienced, which were assumed to be uniformly distributed throughout. 
However, the worm's movements obviously influence the statistical properties of its perceived concentration changes, which can be thought of as a specific instance of a general phenomenon known as ``information self-structuring,'' where an organism actively selects and shapes the sensory inputs that it receives through its actions~\cite{Lungarella2006}. An important direction for future work is to analyze the information flow of a sensorimotor circuit when it is in closed-loop interaction with its environment. 
Finally, to perform our analysis we had access to unlimited, noiseless data. Applying a similar analysis to cell recordings presents a number of challenges with respect to the limits of resolution and noise (for recent work applying information measures to experimental data, see~\cite{Nemenman2004,Paninski2003,Rozell2005,Victor2006,Wibral2014}).

\section*{Methods}
\label{Methods}

\subsection*{Model}

The model used for this study is the same used in our previous study of salt klinotaxis in {\it C. elegans}~\cite{Izquierdo2013}. The model consists of the putative minimal klinotaxis circuit~(Figure 1) connecting the main salt chemosensory class ASE~\cite{Suzuki2008} to the neck motor class involved in modulating the amplitude of the sinusoidal locomotion, SMB~\cite{Gray2005}. 
The circuit was identified by mining the {\it C. elegans} connectome~\cite{White1986} and constraining it using existing experimental and theoretical considerations~\cite{Izquierdo2013}. Chemosensory neurons were modeled after the ON (Eq.~\ref{eq:oncell}) and OFF (Eq.~\ref{eq:offcell}) ASE cells based on observations by Suzuki et al.~\cite{Suzuki2008} and modeled as instantaneous functions of a derivative operator applied to the recent history of attractant concentration (Eq.~\ref{eq:sensor})~\cite{Schnitzer1990}. Interneurons were modeled as passive, isopotential nodes (Eq.~\ref{eq:interneuron}). The model includes chemical synapses and electrical gap junctions. Chemical synapses were modeled as a sigmoidal function of presynaptic voltage, $\sigma(x)=1/(1+e^{-x})$~\cite{Davis1989}. Gap junctions were modeled as a nonrectifying conductance between two cells (penultimate term in Eq.~\ref{eq:interneuron})~\cite{Liu2006}. Neck motor neurons were modeled as passive, isopotential nodes with self-connections representing the voltage dependence of inward currents (Eq.~\ref{eq:motorneuron})~\cite{Goodman1998}, and an additional input from an oscillatory component (Eq.~\ref{eq:pg}). The model worm moves forward in undulatory fashion (Eq.~\ref{eq:neck})~\cite{Gray1964}, driven by the dorsal and ventral motor neurons. The effect of the neck motor neurons on the neck muscles were based on results by Gray et al.~\cite{Gray2005}. The worm is represented as a single point $(x, y)$ with instantaneous velocity, $v$ (Eq.~\ref{eq:body}).  The evolved model employs two kinds of turns to steer towards a source: a sharper and milder head sweep, both of which are consistent with was has been experimentally observed in {\it C. elegans}~\cite{Kim2011}. Altogether, the model is specified by the following set of equations:
	 
\begin{equation}  \label{eq:oncell}
  y_{\text{ON}}(t)=\begin{cases}
    d(t), & \text{if $d(t)>0$}.\\
    0, & \text{otherwise}.
  \end{cases}
\end{equation}

\begin{equation}  \label{eq:offcell}
  y_{\text{OFF}}(t)=\begin{cases}
    0, & \text{if $d(t)>0$}.\\
    -d(t), & \text{otherwise}.
  \end{cases}
\end{equation}

\begin{equation} \label{eq:sensor}
d(t) = \frac{\sum_{t-N}^{t}c(t)}{N}-\frac{\sum_{t-(N+M)}^{t-N}c(t)}{M}
\end{equation}
	 
\begin{equation} \label{eq:interneuron}
\tau_i \frac{d y_i}{d t} = - y_i + \sum_{j=1}^N w_{ji} \sigma(y_j + \theta_j) + \sum_{k=1}^N g_{ki}(y_k - y_i) + I_i
\end{equation}

\begin{equation} \label{eq:motorneuron}
\tau_i \frac{d y_i}{d t} = - y_i + \sum_{j=1}^N w_{ji} \sigma(y_j + \theta_j) + w_{\text{PG}} y_{\text{PG}}
\end{equation}
	 
\begin{equation} \label{eq:pg}
y_{\text{PG}} = \text{sin}(2\pi t / T)
\end{equation}

\begin{equation} \label{eq:neck}
\phi = \frac{d \mu}{d t} = w_{\text{NMJ}}(\sigma(y_{\text{D}}+\theta)-\sigma(y_{\text{V}}+\theta)
\end{equation}

\begin{equation} \label{eq:body}
\vec{v}(t) = \left( \frac{dx}{dt}, \frac{dy}{dt} \right) = \left( v \text{cos}(\mu (t)), v \text{sin}(\mu(t)) \right)
\end{equation}

\noindent where $c(t)$ is the concentration at time $t$; $N$ and $M$ are the durations of the two intervals over which the concentration is averaged such that in response to a step in concentration at time t=0, the sensory cells yield a linear rise to the peak at $t=N$, and a linear decay to base line at $t=N+M$; accordingly, $N$ and $M$ are referred to as the ``rise time'' and ``decay time'' of the sensory neurons; $y$ represents the membrane potential (or neuron activation) relative to the resting potential (thus $y$ can assume positive and negative values); $\tau$ is the time-constant; $\sigma(x)$ is the synaptic potential or output of the neuron; $\theta$ is a bias term that shifts the range of sensitivity of the output function; $w_{ji}$ represents the strength of the chemical synapse; $g_{ki}$ represents the conductance between cell $k$ and $i$ ($g_{ki} >0$); $w_{\text{PG}}$ represents the strength of the connection from the pattern generator; $T$ represents the duration of a one cycle of locomotion on agar ($T = 4.2 \text{sec}$)~\cite{Ferree1999}; $y_{\text{D}}$ and $y_{\text{V}}$  represent the activations of the dorsal and ventral neck motor neurons, respectively; $w_{\text{NMJ}}$ is the strength of the connection from motor neurons to muscles; and $v$ is a constant speed of 0.022 cm/s~\cite{Ferree1999}. 

The nervous system and body model are both idealized. 
First, the number of contacts is not included in the model. Although the anatomical connectivity of the nervous system of the nematode worm {\it C. elegans} has been reconstructed completely~\cite{White1986}, the sign and strength of the anatomical contacts is almost entirely unknown. The sign and strength of the connections in the model were optimized. 
Second, the body model assumes: (a) each body segment follows the one anterior to it; (b) neck muscle length is proportional to motor neuron output; (c) the turning angle is proportional to the difference in muscle length, and (d) the worm can steer by modulating only the amplitude of the sinusoidal movement in the neck angle. 
Finally, although there is debate about the origin of the oscillatory pattern that drives locomotion~\cite{Gjorgjieva2014}, our model is neutral with respect to this debate. The pattern could equally well be generated from proprioceptive feedback through the body and environment during locomotion~\cite{Niebur1993,Bryden2008,Boyle2012} or from a neuron or a network of neurons that produce the pattern centrally~\cite{Niebur1991,Wen2012,Karbowski2008}. The model does assume, however, that the oscillatory input enters the circuit at the neck motor neurons. 

The unknown parameters of the model where evolved using a genetic algorithm (20 parameters in total). To ensure undulatory locomotion, network parameters were constrained to be symmetrical across the dorsal/ventral midline. Fitness was evaluated in simulated chemotaxis assays using a conical concentration gradient (Eq.~\ref{eq:gradient}). At the start of each assay, the model worm was placed with a random orientation at a point 4.5 cm from the peak of the gradient and motor neuron activations were randomized over the range $[0, 1]$. Gradient steepness, $\alpha$, was randomized over the range $[-0.38, -0.01]$. The fitness score was quantified in terms of a chemotaxis index $CI$ defined as the time average of the distance to the peak of the gradient (Eq.~\ref{eq:fitness}), which has also been used to quantified chemotaxis performance experimentally~\cite{PierceShimomura1999}. In preliminary experiments we found that circuits with a chemotaxis index above 0.75 performed highly efficient klinotaxis. We focused only on the highest-performing subpopulation, namely those networks having a chemotaxis index (CI) of at least 0.75 (n = 27). 

\begin{equation} \label{eq:gradient}
c(t)=\alpha \sqrt{x(t)^2 + y(t)^2}
\end{equation}

\begin{equation} \label{eq:fitness}
CI=1 - \int_0^T \frac{c(t)}{c(0)} dt
\end{equation}

\noindent where $c(t)$ is the Euclidean distance to the peak at time $t$, $h(0)$ is the model worm's initial distance from the peak (4.5 cm), and $T$ is the total simulated assay time (500 sec).

\subsection*{Information Theoretic Tools}
The central concept of information theory is {\it entropy}~\cite{Shannon1948,Thomas2006}, a measure of uncertainty about the outcome of a measurement on a random variable $X\in\{x_1, ... , x_N\}$, defined as 
 
\begin{equation} \label{eq:entropy}
H(X)=\sum_i p(x_i)\text{log}\frac{1}{p(x_i)}
\end{equation}

\noindent where $p(x)$ gives the probability of outcome $x$. 

{\it Mutual information} is a measure of the dependence between two random variables, $X$ and $Y$; it quantifies the amount by which a measurement on one of the variables reduces our uncertainty about the other, defined as
 
\begin{equation} \label{eq:mutualinf}
I(X;Y)=\sum_i \sum_j p(x_i,y_j) \text{log} \frac{p(x_i,y_j)}{p(x_i)p(y_j)}
\end{equation}

\noindent where $p(x,y)$ gives the joint probability distribution of $X$ and $Y$.
 
Shannon's information measures average across all measurement outcomes.  In order to perform a more fine-grained analysis of informational relationships, we need to unroll such averages using a measure of specific information~\cite{DeWeese1999,Butts2003}. Unfortunately, there is no general consensus in the literature about the best way to do this. Indeed, an infinite number of measure exist that satisfy the basic requirement of measure of specific information while differing in their details. Accordingly, we have adopted the most commonly used measure of specific information in the literature~\cite{Eckhorn1975,Balduzzi2008,Borst1999},

\begin{equation} \label{eq:mutualinf}
I(X=x_i;Y)=\sum_j p(y_j | x_i) \left[ \text{log} \frac{1}{p(x_i)}  - \text{log} \frac{1}{p(x_i|y_j)}  \right]
\end{equation}

\noindent where $p(y | x)$ gives the conditional probability of $x$ given $y$. 

{\it Transfer entropy} provides a general measure of the influence that one stochastic process has on another~\cite{Schreiber2000}, defined by: 
 \begin{equation} \label{eq:transferentropy}
T_{Y\rightarrow X} = I(X_{t};Y_{t-1} | X_{t-1}) 
\end{equation}
  \noindent where $Y_t$ represents the state of $Y$ at time $t$, and likewise for $X$. Transfer entropy quantifies the information that the previous state of $Y$ provides about the next state of $X$ when conditioned on $X$'s own history.  
Since our model satisfies the Markov property, meaning that its current state is conditionally independent from its history when given knowledge of its previous state, we consider information transfer only from the previous state of $Y$ while conditioning on only the previous state of $X$. For systems that are not known to satisfy the Markov property, including essentially all real-world data, one should also consider information transfer from longer histories of $Y$ while conditioning on longer histories of $X$~\cite{Faes2011,Kaiser2002}.

%{\color{red} Multivariate transfer entropy is also not considered for our system, as every component of our system has exactly two inputs, except the AIY cells. For components with two inputs, since each component is a deterministic function of its inputs we know that the transfer from both inputs must be equal to the total information carried by the destination. For the AIY cells, which have three inputs (from the two sensors and the other AIY cell via a gap junction), we know that two of the inputs (those from the two sensors) provide information about completely different ranges of stimuli, so that any transfer from those sources should combine additively in the destination.}

Standard information measures, and even dynamic information measures like transfer entropy~\cite{Schreiber2000,Kaiser2002,Wibral2014}, are typically treated as atemporal. The aim of this work is to characterize how information is carried by individual components and transferred between components over time, so we unroll these measurements over time to perform a more fine-grained analysis of the temporal structure of information flow~\cite{Williams2010B}. 

The first argument to the mutual and specific information measures ($X$) corresponds to the stimulus feature, which in this paper corresponds to the change in concentration ($\dot{c}$, $\Delta c$). The second argument ($Y$) refers to the various components of the brain-body-environment system: the activation of chemosensory cells ($y_{\text{ON}}$,  $y_{\text{OFF}}$), the potential of each of the cells ($y_e$ for $e \in \{$AIYL, AIYR, AIZL, AIZR, SMBDL, SMBDR, SMBVL, SMBVR$\}$) and the worm's neck angle ($\phi$).

In order to calculate the probability distribution, we first evaluate the circuit's behavior for a sample of the changes in concentration, recording the trajectories of all neural and bodily state variables for each stimulus presentation. From the values taken on by each state variable at each moment in time and the corresponding stimuli that produced them, we estimate a joint probability distribution over values of the state variable and the stimulus feature. The time-varying probability distributions are estimated over a fine grid of possible stimulus values, activation values, and times using a kernel density estimation technique known as average shifted histograms~\cite{Scott1985}. For the analysis we used 2200 stimulus presentations, a grid of 50 bins, and 12 shifts along each dimension. Although bin size can affect the quantities of information, we verified that the results of our analysis are qualitatively robust (i.e., the overall pattern of the flow was preserved, including the relative orderings of magnitudes and the locations of the minima and maxima) over a wide range of grid resolutions (from 20 to 200 bins). 

Since we are concerned here with how information about changes in concentration flows throughout the klinotaxis circuit, all informational quantities are normalized by the entropy of the change in concentration, in order to obtain measures that run from 0 to 1.  A measure of 0 indicates that the stimulus is completely indistinguishable from other stimuli based on knowledge of the observed component, while a measure of 1 indicates that the stimulus can be uniquely determined from knowledge of the observed component.

For the main informational analysis, the sample used to generate the probability distribution for each variable was chosen from a uniform distribution. The distribution used was calibrated to the ranges experienced by the model worm during simulated klinotaxis. A uniform distribution is a sensible initial choice to understand the circuit outside of the context of the task. It is also the distribution that has been applied experimentally to the study of the chemosensory neurons in the circuit~\cite{Suzuki2008}. In order to study the flow of information in the circuit in relation to the specific distribution of stimulus that the model receives during klinotaxis behavior, we repeated the analysis using the empirical distribution generated from repeated simulations (shown in Figure 9D).

In terms of notation, given the feedforward structure of the putative minimal klinotaxis circuit, we grouped the information flow analysis by neuron class: ASE, AIY, AIZ, and SMB. In order to make the notation more readable, we consider the joint information between two neurons in the same class by their class name. Thus, ASE represents the joint set of variables \{ASEL, ASER\}, AIY represents \{AIYL, AIYR\}, AIZ represents \{AIZL, AIZR\}, and SMB represents the set \{SMBDL, SMBDR, SMBVL, SMBVR\}. To represent different subsets of SMB we also used a shortened notation. For example, SMBV represents the set \{SMBVL, SMBVR\} and SMBD represents \{SMBDL, SMBDR\}.

\subsection*{Assays}
We analyzed the network under two different conditions: a concentration step assay and an information clamp assay. The concentration step assay involves giving the circuit a step in concentration at a specific time, a procedure analogous to that used experimentally (e.g.,~\cite{Goodman1998,Suzuki2008,Thiele2009,Oda2011}). In our analysis, the magnitude of the step ranged uniformly between $\pm0.01$, which is the approximate range of concentration changes experienced during a typical simulated klinotaxis run. 

In order to analyze the components with time-dependent input (i.e., SMB and neck angle), we developed a second assay that involves giving the circuit a constant change in concentration over time. As the sensory neurons detect changes in concentration over time, a constant change in concentration produces no change in the activity of the chemosensory neurons, acting as an information clamp for the components in the network that do not depend on the cyclic input (ASE, AIY, AIZ). Needless to say, neither of the two assays perfectly emulates what the actual worm experiences during a klinotaxis run. They are simplifications that allow us to study the properties of the circuit under idealized circumstances. 

%\section*{Acknowledgments}
%This work was supported in part by NSF grant IIC-1216739.

%\section*{References}
% The bibtex filename
\bibliography{bib}

\newpage 
\section*{Supporting Information }

\begin{figure}[ht!]
\begin{center}
\includegraphics[width=0.4\textwidth]{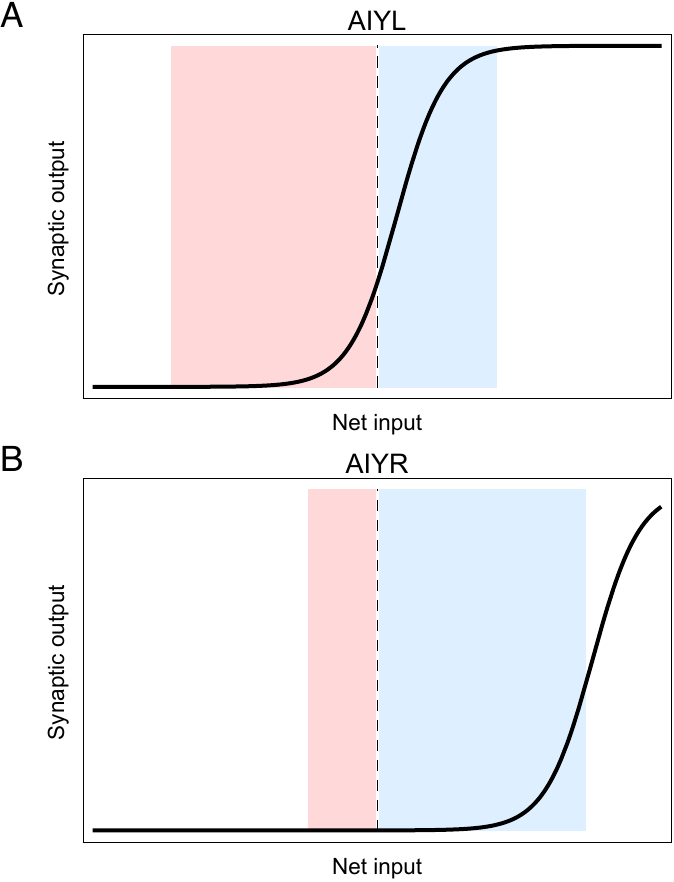}
\caption*{
\noindent {\bf 
Figure S1.
Mechanism of information asymmetry in AIY.
}
Synaptic transfer functions for the left (A) and right (B) AIY cells (solid black).
Resting potential of the cells shown with dashed black line. ASER connects to both AIY cells through an inhibitory chemical synapse. ASEL connects to both AIY cells through an excitatory chemical synapse. Activity in ASER/ASEL drives the inputs to both AIY cells into the red/blue region, respectively. The relative strength of the connections is shown by the size of the region. The responsiveness of AIYL is the result of the alignment between the sensitive area of the synaptic transfer function and the range of possible net input. The bias in AIYR shifts the synaptic transfer function, leaving the cell sensitive only to the largest positive changes in concentration. Changes in the activation in AIYL transmitted through the gap junction are equally ineffective to AIYR due to the shifted sensitive region.
}
\label{SuppFig1}
\end{center}
\end{figure}

\vspace{5 mm}

\begin{figure}[ht!]
\begin{center}
\includegraphics[width=0.4\textwidth]{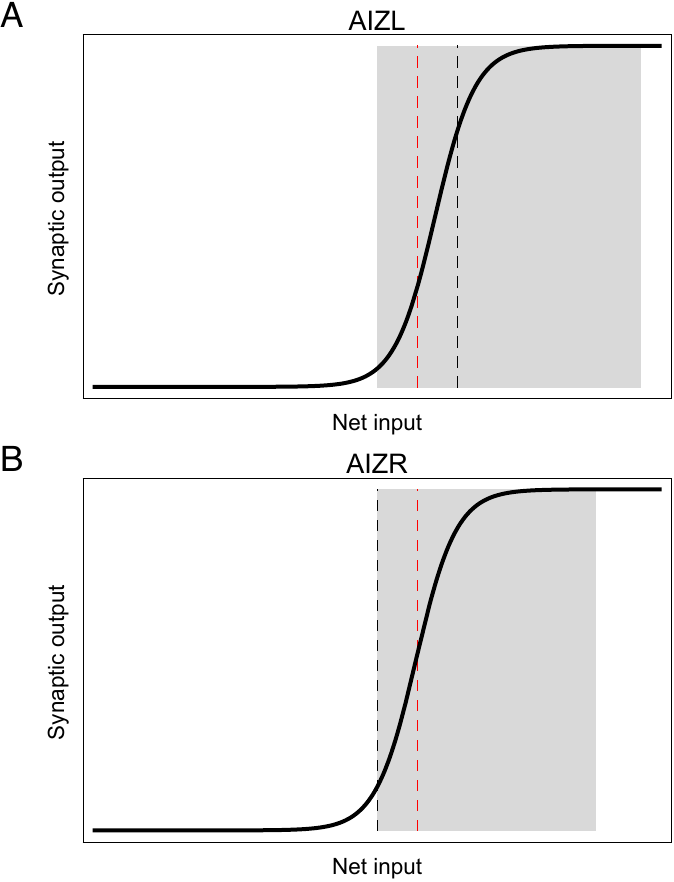}
\caption*{
\noindent {\bf 
Figure S2.
Mechanism of information symmetry in AIZ.
}
Synaptic transfer functions for the left (A) and right (B) AIZ cells (solid black).
Resting potential of the cells before effects from the gap junction shown with dashed black line. Resting potential of the cells after equalization from the gap junction exchange shown with red dashed line. AIZ cells have incoming excitatory chemical synapses from AIY cells, left and right respectively. Therefore, activity in AIYR/AIYL drives the inputs to AIZR/AIZL into the gray region, respectively. The relative strength of the connections are shown by the size of the region. However, because AIYR shows very little activity, changes in activation in AIZR are not due to the chemical synapse; instead they are due to changes in the level of activation of AIZL through the gap junction. Unlike in AIY, the balance of the resting potential and the sensitive region of the synaptic transfer function in both AIZ cells results in a response to changes in concentration to negative and positive changes in concentration.
}
\label{SuppFig2}
\end{center}
\end{figure}

\vspace{5 mm}

\begin{figure}[ht!]
\begin{center}
\includegraphics[width=0.4\textwidth]{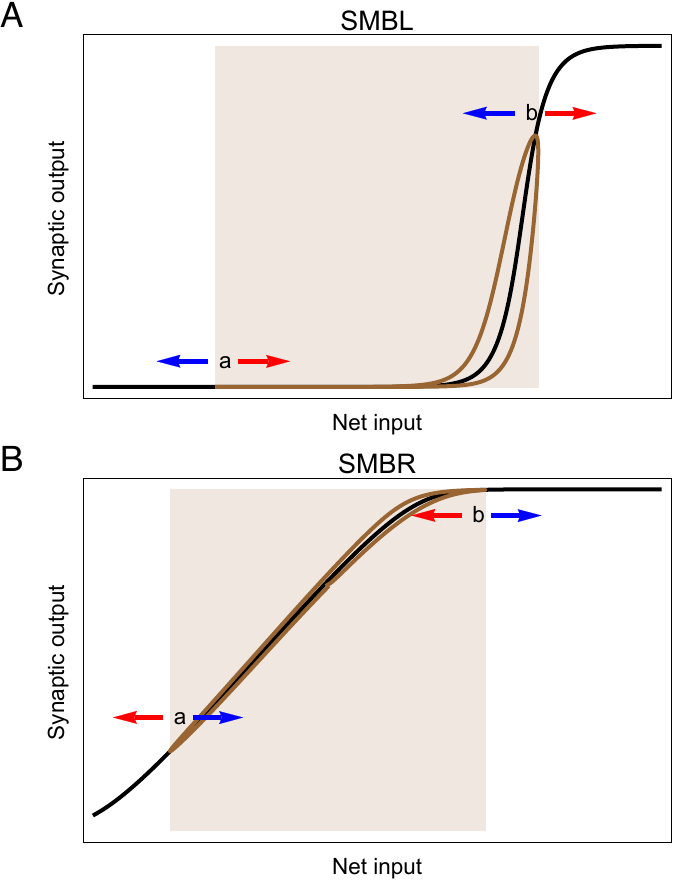}
\caption*{
\noindent {\bf 
Figure S3.
Mechanism of information gating in SMB.
} 
Synaptic transfer functions for the left (A) and right (B) pair of dorsal and ventral motor neurons, SMBL and SMBR, respectively (black trace). Instantaneous synaptic output as a function of net input when the head sweep oscillation is present (brown trace). Shaded areas show the range of oscillation due to the incoming connections from the pattern generator. For each of the SMB pairs, the input to the dorsal and ventral cells moves out of phase over the brown trajectory. As a result, when the dorsal motor neuron is at point $a$ in the curve, the ventral motor neuron is at point $b$, and vice versa.  Red and blue arrows show the effects of negative and positive changes in concentration on the input to the motor neurons, respectively.  Due to the shift in the biases for the synaptic transfer functions, a change in concentration sometimes results in a change in the dorsal but not the ventral synaptic output, and viceversa. For the SMBL pair (A), a change in concentration results in a change in the synaptic output of the neuron in $b$, but not of the neuron in $a$.  For the dorsal/ventral SMBR pair (B), $a$ and $b$ represent the opposite regions: the neuron at $a$ is more sensitive to changes in input than the other neuron at $d$.  To different degrees, the same is the case for other points along the curve. 
In both pairs of SMB neurons, the result is an antiphase dorsal/ventral gating mechanism. 
}
\label{SuppFig3}
\end{center}
\end{figure}

\vspace{5 mm}

\begin{figure}[ht!]
\begin{center}
\includegraphics[width=0.9\textwidth]{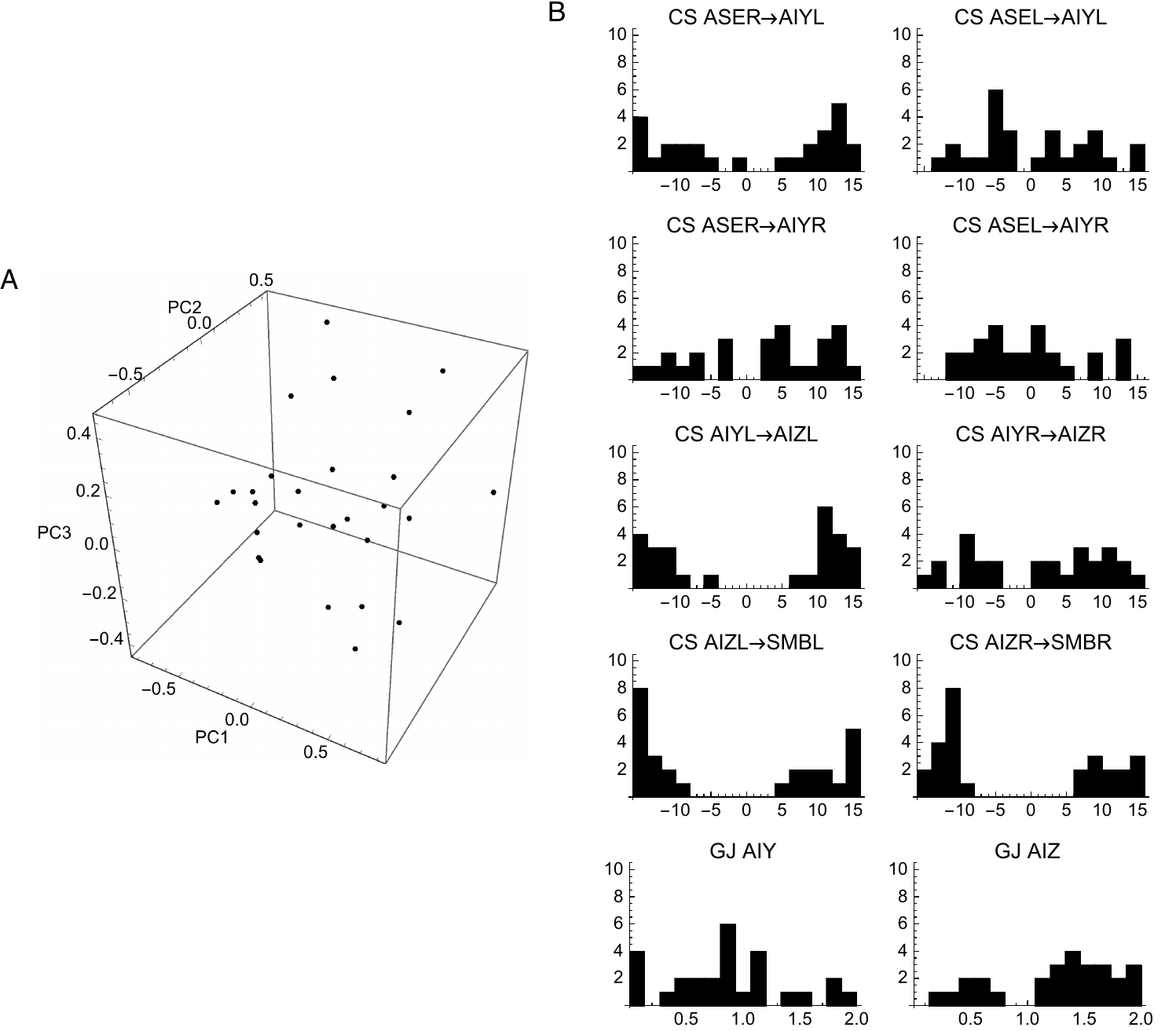}
\caption*{
\noindent {\bf 
Figure S4. 
Cellular and synaptic properties of the ensemble of successful klinotaxis networks.} (A) Principal component projection. The first three principal components capture 52.1\% of the variance in the ensemble. There is no clear clustering of networks in this space.  (B) Distribution of the chemical synapses (CS) and gap junctions (GJ) in the ensemble. Most connection strengths vary over the allowable ranges.
}
\label{SuppFig4}
\end{center}
\end{figure}

\end{document}